\documentclass{jfm}
\usepackage{graphicx}
\usepackage{natbib}
\usepackage{amsmath}
\usepackage{upgreek}
\usepackage{epstopdf, epsfig}
\usepackage{color}
\usepackage{soul}

\newcommand{\vect}[1]{\boldsymbol{#1}}

\shorttitle{Waves in screeching jets}
\shortauthor{Edgington-Mitchell et al.}

\title{Waves in screeching jets}

\author{Daniel Edgington-Mitchell\aff{1}
  \corresp{\email{daniel.mitchell@monash.edu}},
 Tianye Wang\aff{2},
  Petronio Nogueira\aff{1},
 Oliver Schmidt\aff{3},
  Vincent Jaunet\aff{4},
   Daniel Duke\aff{1},
Peter Jordan\aff{4},
 \and Aaron Towne\aff{2}}

 \affiliation{$^1$Dept. of Mechanical and Aerospace Engineering, Monash University,
Melbourne, VIC 3800, Australia\\[\affilskip]}
 
\affiliation{\aff{1}Dept. of Mechanical and Aerospace Engineering, Monash University, VIC 3800, Australia\\
\aff{2}Department of Mechanical Engineering, University of Michigan, USA.\\
\aff{3}Department of Mechanical and Aerospace Engineering, University of California San Diego, USA\\
\aff{4}Dept. Fluides, Thermique, Combustion, Institut PPRIME, CNRS - Universite de Poitiers, ENSMA, UPR 3346, 86036 Poitiers, France}

\pubyear{1}
\volume{1}
\pagerange{1}
\date{?; revised ?; accepted ?. - To be entered by editorial office}
\begin{document}

\maketitle

\begin{abstract}
The interaction between various wavelike structures in screeching jets is considered via both experimental measurements and linear stability theory. Velocity snapshots of screeching jets are used to produce a reduced order model of the screech cycle via proper orthogonal decomposition. Streamwise Fourier filtering is then applied to isolate the negative and positive wavenumber components, which for the waves of interest in this jet correspond to upstream and downstream-travelling waves. A global stability analysis on an experimentally derived base flow is conducted, demonstrating a close match to the results obtained via experiment, indicating that the mechanisms considered here are well represented in a linear framework. In both the global stability analysis and the experimental decomposition, three distinct wavelike structures are evident. These three waves are those first shown by \cite{TamHu} to be supported by a cylindrical vortex sheet. One of these is the well-known downstream-travelling Kelvin-Helmholtz mode. Another is the upstream-travelling guided acoustic jet mode that has been a topic of recent discussion by a number of authors. The third component, with positive phase velocity, has not previously been identified in screeching jets. Via a local stability analysis, we provide evidence that this downstream-travelling wave is a duct-like mode similar to that recently identified in high-subsonic jets by \cite{TowneetalJFM2017}. We further demonstrate that both of the latter two waves are generated by the interaction between the Kelvin-Helmholtz wavepacket and the shock cells in the flow, according to a theory first proposed in \cite{tam1982shock}. Finally, we consider the periodic spatial modulation of the coherent velocity fluctuation evident in screeching jets, and show that this modulation is the result of the superposition of the three wavelike structures, with no evidence that the shocks in the flow modulate the growth of the Kelvin-Helmholtz wavepacket. 
\end{abstract}


\maketitle

\section{Introduction}
Shock-containing free shear flows frequently exhibit some form of aeroacoustic resonance, the best-known of which are those that produce screech or impingement tones \citep{edgington2019aeroacoustic}. These resonance mechanisms can be divided into four discrete processes: a downstream-travelling wave \citep{tam1990theoretical,gudmundsson2011instability,sinha2014wavepacket}, a downstream-reflection mechanism  \citep{ManningLele2000AIAA,SuzukiLele,shariff2013ray,berland2007numerical,edgington2018sound}, an upstream-travelling wave \citep{TamHu,ShenTam,bogey2017feedback,gojon2018oscillation,edgington2018upstream,jordan2018jet}, and an upstream-reflection or receptivity mechanism in the nozzle plane \citep{barone2005receptivity,mitchell2012visualization,weightman2019nozzle,karami2020receptivity}. Of these, the downstream-travelling wave is generally thought to be the only process where energy is provided to the resonance loop \citep{tam1990theoretical}; the growth of the instability wave is driven by the extraction of energy from the mean flow. Amplitude prediction models for aeroacoustic resonance have remained elusive, and while there are several models capable of frequency prediction \citep{PowellScreech1953,tam1986proposed}, these models often struggle to fully explain the staging behaviour typical of jet screech \citep{mancinelli2019screech}. In this context, staging behaviour refers to the tendency of resonant systems to experience discontinuous changes in tone frequency with small changes in operating conditions. In screeching axisymmetric jets, these stages are typically classified into A1 \& A2 ($m=0$), B \& D (flapping), and C ($m=1$) helical modes.

\subsection{Upstream-travelling waves in jet resonance}
Screech, like other resonant processes in jets, involves an energy exchange between upstream- and downstream-travelling waves. It is generally accepted that the downstream-travelling wave relevant to resonance in jet screech is the Kelvin-Helmholtz (KH) wavepacket. The nature of the upstream-travelling wave is less clear. \cite{PowellScreech1953} originally conceived the upstream-travelling wave as a freestream acoustic wave, a view that went unchallenged for many decades. The supporting evidence for this theory was quite strong: a sharp tone is evident in the farfield acoustics, and resonance models based on an upstream-travelling wave with sonic phase speed generally predicted frequency well. It was only in the work of \cite{ShenTam} that an alternative was proposed: that the upstream-travelling wave is not a freestream acoustic wave, but rather a guided jet mode. Evidence for the role of this wave was first provided for subsonic impinging jets \citep{tam1990theoretical}, then supersonic impinging jets \citep{bogey2017feedback}, and finally jet screech \citep{gojon2018oscillation,edgington2018upstream}. Frequency prediction models based on the upstream-travelling guided jet mode outperform those that assume a freestream acoustic wave, at least for the $m=0$ screech modes \citep{mancinelli2019screech,mancinelli2019reflection}. It should be noted however that at this point there is still evidence that some resonant process are indeed closed by freestream acoustic waves; \cite{weightman2019nozzle} provide evidence for such closures in various supersonic jet impingement cases. 

\subsection{Other waves present in supersonic jets}
The guided acoustic mode is one of three families of waves that can be supported by a supersonic jet. These waves were first visualized by \cite{oertel1980mach}. \cite{TamHu} then demonstrated that an inviscid high-speed jet can support three families of waves: the Kelvin-Helmholtz wave, subsonic instability waves, and supersonic instability waves. The supersonic instability waves are only present for very high jet Mach numbers, whereas the subsonic waves are present across all Mach numbers. For supersonic jets, these subsonic instability waves, which can propagate both upstream and downstream, are organised hierarchically according to their azimuthal and radial order. With the exception of the axisymmetric mode of radial order 1, the upstream-travelling waves are confined to a narrow frequency band, whereas the downstream-travelling wave can be supported across a wide range of frequencies. The upstream-traveling wave identified in the vortex-sheet dispersion relation is the guided acoustic mode that has been demonstrated to play a significant role in resonance, including the global instability of hot jets and wakes \citep{martini2019acoustic}. The downstream-travelling waves have not previously been discussed in the context of supersonic resonance, but they have been discussed extensively for high Mach-number subsonic jets in the works of \cite{TowneetalJFM2017}, \cite{schmidt2017wavepackets}, and \cite{jordan2018jet}. Though both the upstream- and downstream-travelling waves are associated with the same fundamental mechanism, they have distinct radial structures. In supersonic jets, the upstream-travelling wave has support outside the shear layer of the jet, a requirement for an upstream-travelling wave in a flow that is travelling downstream at supersonic velocity. The downstream-travelling wave by contrast remains essentially trapped within the core of the jet; the work of \cite{TowneetalJFM2017} demonstrated that this mode also obeys the dispersion relation for a soft-walled duct, essentially treating the shear layer of the jet as a pressure-release boundary.  While these downstream-traveling duct-like modes have strictly negative phase velocities in subsonic jets, they can have either negative or positive phase velocity in supersonic jets \citep{Towne2019investigation}.

\subsection{Wave interactions in screeching jets}
It is generally accepted that screech tones are produced by some interaction between the KH waves and the shock/expansion structure in the jet core. The first model for the prediction of screech frequency was that of \cite{PowellScreech1953}. Powell assumed that interactions between the downstream-travelling wave and the shocks could be modelled as emission from a phased-array of equispaced monopoles located at the shock reflection points. Observing that screech radiates most strongly in the upstream direction, Powell further assumed maximum upstream directivity as a requirement for screech (i.e. that waves from all three sources would arrive at the nozzle simultaneously and thus provide constructive reinforcement), and on this basis produced his predictive equation for screech,
\begin{center}
\begin{equation}
\label{eq:Powell}
f=\frac{U_{c}}{s(1+M_{c})}.
\end{equation}%
\end{center}%

Here $f$ is the frequency of the screech tone, $s$ is the spacing of the shock cells, and $U_c$ and $M_c$ are the convection velocity and convective Mach number respectively. In a later paper \citep{powell1992observations}, Powell reconsidered this model and stated that there was no reason to assume perfect reinforcement at the nozzle lip was a requirement for screech. Nonetheless this model does an admirable job of predicting screech across a range of operating conditions, but is incapable of accounting for the mode-staging behaviour typical of aeroacoustic resonance.

An alternative model was proposed by \cite{tam1986proposed}, an extension of work begun in \cite{tam1982shock}. In the earlier work, Tam and Tanna construct a model for broadband shock noise, on the assumption of weak interaction between travelling KH waves and stationary shock waves. Key points of the model are recapitulated here, though the nomenclature used is slightly different than in the original paper. The downstream-travelling KH waves can be modelled as a wavepacket of the form

\begin{center}
\begin{equation}
\label{eq:TT1}
u_\textit{kh}=a(x)\psi(r)e^{i(k_\textit{kh}x-\omega t)}.
\end{equation}%
\end{center}%

Here $u_\textit{kh}$ represents a velocity perturbation associated with the KH wavepacket, $a(x)$ is a spatial amplitude distribution, $\psi(r)$ is the radial eigenfunction of the KH wavepacket, $k_\textit{kh}$ is the wavenumber and $\omega$ the frequency.

The spatial modulation of velocity ($u_s$) by the quasi-stationary shock cell structures within the flow was modelled in the work of \cite{tam1982shock}  using the vortex-sheet approach of \cite{prandtl1904stationaren} and \cite{pack1950note}, which can be expressed in simplified form as

\begin{center}
\begin{equation}
\begin{gathered}
\label{eq:PP}
u_s=\sum_{n=1}^{\infty} A_n (e^{i k_{s_n} x}+e^{-i k_{s_n} x})\\
\end{gathered}
\end{equation}%
\end{center}%

Here $A_n$ defines the amplitude of each shock cell mode, while $k_{s_n}$ defines the wavenumber, for $n = 1,2,3\dots$. As stated in \cite{tam1982shock} and shown more explicitly in \cite{RayLele}, the interaction between the KH wavepacket and the stationary shocks can be represented by the product of the two wave expressions. Ignoring amplitude terms, for the first shock cell mode $n=1$ this can be written as 
\begin{center}
\begin{equation}
\label{eq:TT2}u_\textit{kh} u_s\propto e^{(i(k_\textit{kh}+k_s) x -i \omega t)}+e^{(i(k_\textit{kh} - k_s) x -i \omega t)}.
\end{equation}%
\end{center}%


Alternatively, this relation can be directly obtained from the analysis a convective term of the Navier-Stokes equations. Considering only the first shock-cell mode, the mean flow in the streamwise direction $U$ can be written as

\begin{equation}
    U(x,r)= U_{sm}(x,r) + U_{sh}(x,r) \frac{1}{2} \left( e^{i k_{s} x} + e^{-i k_{s} x} \right),
    \label{eqn:Ushocks}
\end{equation}

\noindent where $U_{sm}$ is the shock-less mean flow (which can be obtained using a low-pass filter, for example), $U_{sh}$ is the slow-varying part of the shock-cell structure (or the envelope), which includes the amplitude term $A_1$, and $k_{s}$ is the wavenumber associated with its oscillatory part. The subscripts $sm$ and $sh$ are used to denote the smooth and shock-related parts of the mean. Expanding the velocity as $\tilde{u}(x,r)=U(x,r)+u(x,r,\theta,t)$, all convective terms of the streamwise momentum equations will have a dependency on both $U_{sm}$ and $U_{sh}$. For instance, the linearised form of $\tilde{u}\partial_x \tilde{u} - U\partial_x U$ is given by

\begin{eqnarray}
    \tilde{u}\partial_x \tilde{u} - U\partial_x U \approx 
    [U_{sm} + U_{sh}\frac{1}{2} \left( e^{i k_{s} x} + e^{-i k_{s} x} \right)] \partial_x u + \nonumber \\
    u \partial_x [U_{sm} + U_{sh}\frac{1}{2} \left( e^{i k_{s} x} + e^{-i k_{s} x} \right)] \\
    = U_{sm} \partial_x u + U_{sh}\frac{1}{2} \left( e^{i k_{s} x} + e^{-i k_{s} x} \right) \partial_x u + u \partial_x U_{sm} + \nonumber  \\
    u (\partial_x U_{sh})\frac{1}{2} \left( e^{i k_{s} x} + e^{-i k_{s} x} \right) + u U_{sh} \frac{1}{2} \left( i k_{s} e^{i k_{s} x} -i k_{s} e^{-i k_{s} x} \right).
   \label{eqn:UshocksLin1}
\end{eqnarray}

Considering that $U_{sm}$ and $U_{sh}$ are slow varying, the streamwise derivatives of these quantities will be disregarded, as in locally parallel analyses. Also, using the normal modes ansatz, disturbances can be written as $u(x,r,\theta,t)=u'(r) e^{-i \omega t + i k_x x + i m \theta}$, where $\omega$ is the frequency, $k_x$ is the streamwise wavenumber and $m$ is the azimuthal wavenumber. Thus, (\ref{eqn:UshocksLin1}) can be rewritten as

\begin{eqnarray}
    \tilde{u}\partial_x \tilde{u} - U\partial_x U \approx 
    i k U u' e^{-i \omega t + i k_x x + i m \theta} \nonumber +\\ 
    + i u' U_{sh} \frac{1}{2} \left( k_{s} e^{-i \omega t+i (k_x+k_{s}) x + i m \theta} -k_{s} e^{-i \omega t+i (k_x-k_{s}) x + i m \theta} \right).
   \label{eqn:UshocksLin2}
\end{eqnarray}

A clear connection between the most amplified structures in the flow and the generation of waves at other wavenumbers is evident in (\ref{eqn:UshocksLin2}). If the wavepacket wavenumber is considered ($k_x=k_{kh}$), (\ref{eqn:UshocksLin2}) becomes equivalent to the well-known expression originally presented by \cite{tam1982shock}, demonstrating that new wavenumbers are energised by the interaction between shocks and the wavepacket.

It is clear from (\ref{eq:TT2}) and (\ref{eqn:UshocksLin2}) that the interaction of the KH wave with the stationary shocks produces two wave-like disturbances for a given frequency. These wavelike disturbances have wavenumbers dictated by the sum and difference of wavenumbers associated with the KH wavepacket ($k_\textit{kh}$) and the shock cells ($k_s$). For the case where the KH wavenumber is smaller than the shock spacing (as is typically observed), the difference wavenumber is negative, and the wave has negative phase velocity. The sum term will always represent a wave that has positive phase velocity. In this work we do not explicitly use the model of \cite{tam1982shock}, however (\ref{eq:TT2}) will be used to explain phenomena observed in Section 3.

The model of \cite{tam1982shock} was originally developed to explain broadband shock-associated noise (BBSAN), but \cite{tam1986proposed} suggested that screech could simply be considered a special case of BBSAN. On this basis the authors developed a predictive model for screech frequency, taking the limit of the BBSAN model of \cite{tam1982shock} as the observer angle approaches the upstream axis,

\begin{center}
\begin{equation}
\label{eq:WL}
f_{s}=\frac{U_{c}k_s}{2\pi(1+M_{c})}.
\end{equation}%
\end{center}%

While this relation provides identical predictions to the model of 
\cite{PowellScreech1953}, its provenance is rather different. The central thesis of \cite{tam1986proposed} is that the screech frequency is selected by the weak interaction of the KH wavepackets with the shock structures; this interaction only produces radiation back to the nozzle lip in a narrow band of frequencies.

\subsection{Wave modulation}
Most existing models for screech assume that the screech tone is produced by some form of interaction between the KH wave and the shock structures within the jet. It is also typically assumed that the KH wavepacket is essentially unaffected by this interaction. Part of the motivation of this paper is the evaluation of the validity of this assumption. It is well recognized that turbulence undergoes significant changes during passage through a shock wave. Both the shock and the turbulence are influenced by this interaction; the shock becomes locally distorted, while the turbulence sees an amplification of intensity and Reynolds stress \citep{ducros1999large}. This amplification has been shown to depend on the scale of the turbulence, with finer scales amplified more than large scales. The interaction of a vortex with a shock strongly depends on the strength of both the shock and the vortex. Passage through a strong shock has been shown to significantly deform the shape of an isolated vortex \citep{grasso2000shock}. The situation in a screeching jet is more complicated: the train of vortices that comprise the wavepacket typically span the sonic line of the jet, meaning some portion of the vortices may pass through the shock cell, whereas components further from the centreline do not. It is thus presently unclear whether the structure or growth of the KH wavepacket is altered or modulated via interaction with the shock cells of the jet.

It is well established that the turbulence fluctuations in the nearfield of the jet are strongly modulated, due to the presence of the standing wave in the acoustic nearfield of the jet \citep{westley1975near,PandaJFM1999}. Formed by the interaction of the downstream-travelling hydrodynamic waves and the upstream-travelling waves, this standing wave is clearly evident in measures of both fluctuating pressure and velocity. Thus measurements of velocity in screeching jets show strong modulation in the axial direction, but the presence of the standing wave makes it difficult to determine whether this is simply the signature of the standing wave, or the shocks modulating the growth of the KH wavepacket.

\subsection{Linear models for the screech problem}
In general, the initial growth of the KH wavepacket can be well predicted by the careful application of linear stability theory \citep{michalke1984survey,morris2010instability}, even for highly turbulent jets \citep{cavalieri2013wavepackets}. One outstanding question is how well such linear models perform in shock-containing flows, and whether or not they can provide a description of the non-linear KH-shock interaction when the shock structure is included in the mean flow. There have been limited attempts to apply linear stability theory to shock containing jets. In a global stability analysis of a shock-free jet, \cite{nichols2011global} observed the subsonic modes of \cite{TamHu}, and suggested that the  upstream-travelling subsonic modes could underpin resonance in shock-containing flows. \cite{beneddine2015global} conducted a global analysis on a laminar, two-dimensional shock-containing jet, and demonstrated that it exhibits a global instability that matches many of the characteristics of jet screech. \cite{beneddine2015global} were also able to demonstrate the sensitivity of this instability to the thickness of the nozzle lip, which has been observed experimentally for both screeching \citep{RamanCess} and impinging \citep{weightman2019nozzle} jets. Nonetheless, the validity of linear models for the screech process remains unclear; there are many processes involved in jet screech that are non-linear.

\subsection{Summary}
There remain a broad range of open questions regarding jet screech. What mechanism selects the screech frequency? Which of the processes underpinning screech can be modelled using linear theory? Does the interaction between the KH wavepacket and the shocks affect the growth of the wavepacket? In an attempt to answer these questions, we pair an extensive experimental database with a range of stability analyses. Experimentally, we consider three jets undergoing screech, two jets characterized by an $m = 0$ instability mode, but with shocks of different strengths, and one jet whose screech is characterized by an $m = 1$ helical mode, with much stronger shocks. The velocity fields recorded for these jets are decomposed on both an energy and spatial wavenumber basis. A global stability analysis is performed on the experimentally-determined baseflow for the case with the weakest shocks. Local stability analysis is used to interpret some of the results from both the experiment and the global analysis.

\section{Database and Methodology}

\subsection{Experimental Database}
\begin{figure}
\centering
\includegraphics[width=1\textwidth]{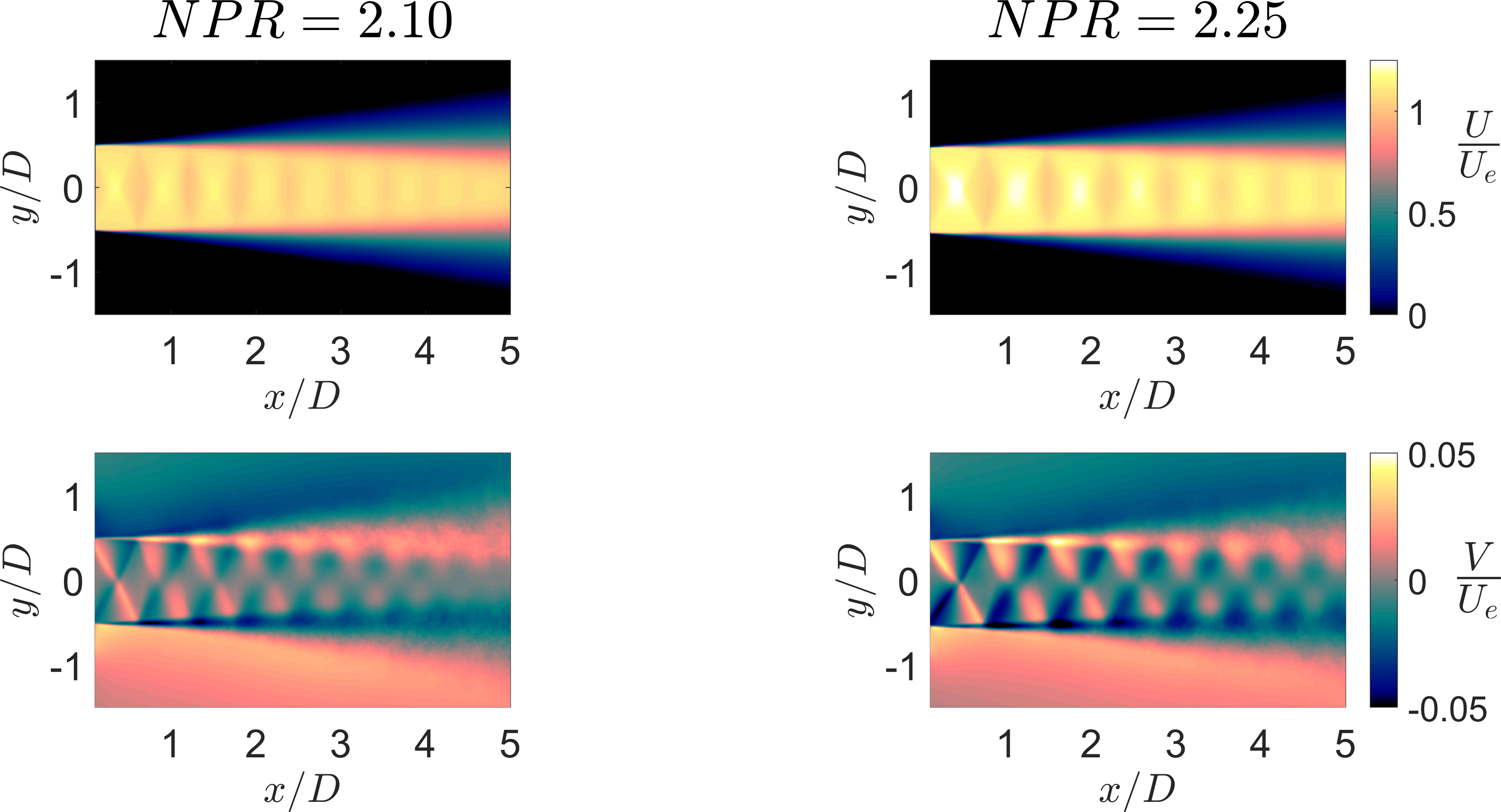}
\caption{Mean velocity fields for the two jets characterized by an $m=0$ screech tone, operating at $\textit{NPR} = 2.10$ and $\textit{NPR} = 2.25$.}
\label{fig:Mean2}
\end{figure}

\begin{figure}
\centering
\includegraphics[width=1\textwidth]{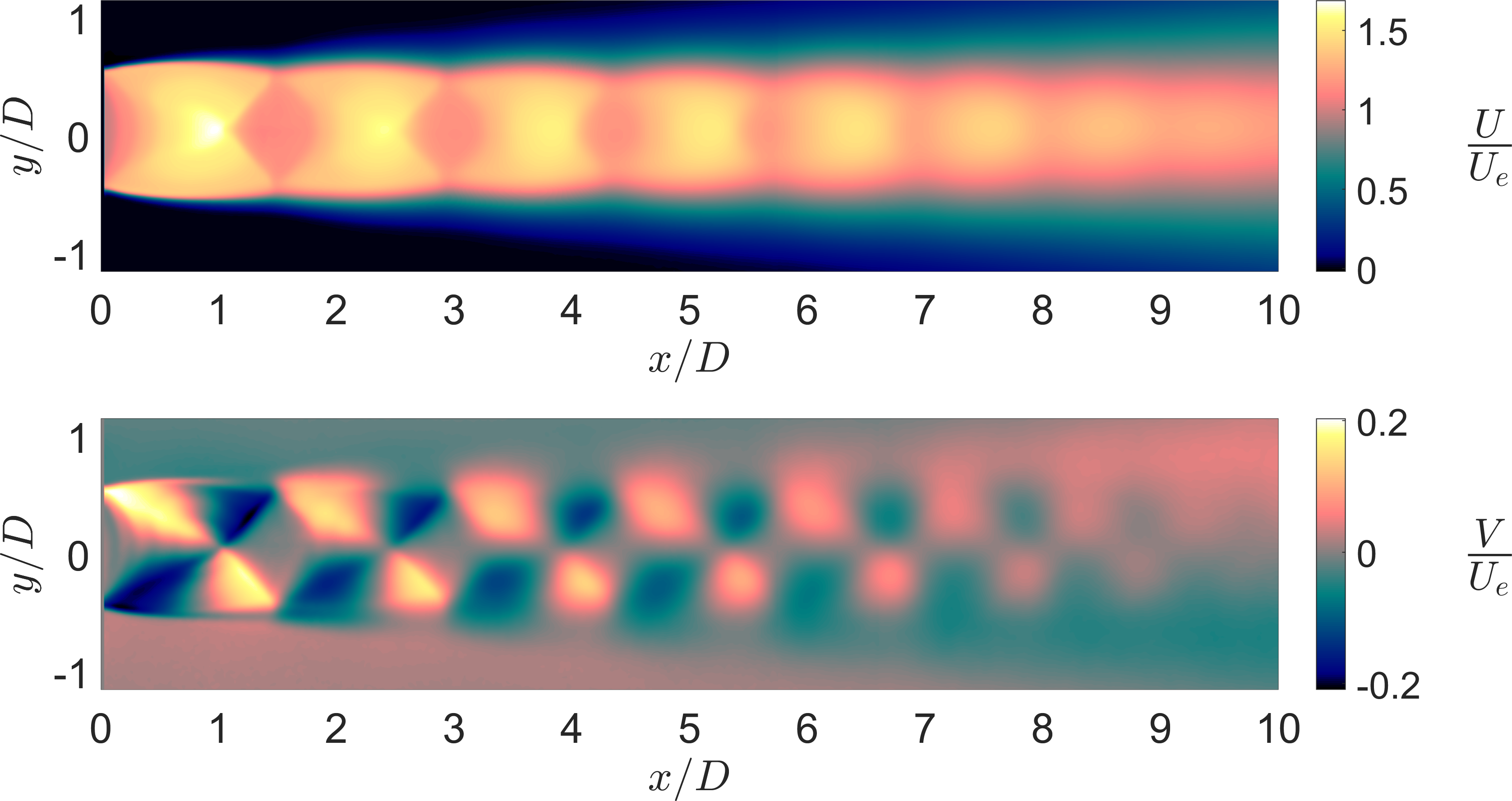}
\caption{Mean velocity fields for the jet characterized by an $m=1$ screech tone, operating at $\textit{NPR} = 3.40$.}
\label{fig:Mean3}
\end{figure}

The experimental database used here has been well documented in prior literature. All cases considered are from similar experimental facilities (and the same nozzle), however the details of the velocimetry differ somewhat. Two cases are presented for the $m=0$ mode, the A1 and A2 modes of jet screech at pressure ratios of $\textit{NPR} = 2.10\,\&\,2.25$ respectively, previously studied in \cite{edgington2018upstream}. Note a slight correction to the Strouhal numbers associated with this data presented in the prior paper. An additional case at a nozzle pressure ratio of $\textit{NPR} = 3.40$ is presented, where the flow is characterized by an $m=1$ mode associated with the helical C screech mode \citep{edgington2014coherent,tan2017novel,li2019shock}. A summary of the relevant parameters for the jets is presented in table \ref{tab:Exp_Param}; $NPR$ refers to ratio of stagnation to ambient pressure, $M_j$, $U_j$ and $D_j$ refer to the ideally-expanded Mach number, velocity and diameter respectively, and the frequency is non-dimensionalized such that $St = fD_j/U_j$. All jets considered here issue from a purely converging nozzle of diameter $D=15$ mm, with a radius of curvature of $67.15$ mm, ending with a parallel section at the nozzle exit, and an external lip thickness of $5$ mm. The nozzles are connected to a large plenum chamber; the area ratio between the nozzle and plenum is approximately 100:1. As a consequence of this high contraction ratio, it is expected that the boundary layer at the nozzle exit will be laminar and extremely thin (below the measurement resolution of the PIV system).

\begin{table}
\centering
\caption{Jet Conditions}
\label{tab:Exp_Param}       
\begin{tabular}{ccccc}
\hline\noalign{\smallskip}
$\textit{NPR}$ &  $M_j$    &   $Re$ & $St$ & Mode  \\
\noalign{\smallskip}\hline\noalign{\smallskip}
2.10 & 1.09 & $4.4 \times 10^5$ & 0.67 & A1 \\
2.25 & 1.14 & $4.7 \times 10^5$ & 0.65 & A2 \\
3.40 & 1.45 & $8.6 \times 10^5$ & 0.26 & C\\
\noalign{\smallskip}\hline
\end{tabular}
\end{table}

For the $\textit{NPR} = 2.10\,\&\,2.25$ datasets, particle images were obtained using a 12-bit Imperx B4820 camera, with a CCD array of $4872 \times 3248$ px, at an acquisition frequency of $2$ Hz. Illumination was provided by a Nd:YAG laser, producing a pair of 6 ns pulses of approximately 160 mJ, separated by $\Delta  t=1$ $\upmu$s. For the $\textit{NPR} = 3.40$ dataset, particle images were obtained using a pair of PCO 4000 cameras mounted orthogonal to the jet, each with a CCD array of $ 4008 \times 2760$ px. The resultant velocity fields from the two cameras were stitched together using using a convolution with an adaptive Gaussian window \citep{agui1987performance} using an overlap of 7.5\%. Illumination was provided by a Nd:YAG laser, producing a pair of 6 ns pulses of approximately 120 mJ, separated by $\Delta  t=0.8$ $\upmu$s.

Both jets were seeded with smoke particles, whose diameter was estimated at 600 nm based on observed relaxation times across a normal shock \citep{mitchell2013near}.  The pertinent PIV parameters are summarized in table \ref{tab:PIV_Param}. The images were analyzed using a multi-grid cross-correlation algorithm \citep{Soria:96}, where $IW_0$ and $IW_1$ refer to initial and final interrogation windows respectively.

\begin{table}
\centering
\caption{Non-dimensional PIV Parameters}
\label{tab:PIV_Param}       
\begin{tabular}{ccc}
\hline\noalign{\smallskip}
Parameter &  Value for \textit{NPR} = 2.1 or 2.25   &   Value for \textit{NPR} = 3.4   \\
\noalign{\smallskip}\hline\noalign{\smallskip}
IW$_0$ & 0.12D & 0.10D   \\
IW$_1$ & 0.030D & 0.026D \\
Grid Spacing $\Delta x$ & 0.01D & 0.013D \\
Depth of Field & 0.04D & 0.17D \\
Light Sheet Thickness & 0.1D & 0.1D \\
Field of View & 5.7D x 3.8D & 10D x 2.2 D \\
Velocity snapshots & 9,000 & 8,000 \\
\noalign{\smallskip}\hline
\end{tabular}
\end{table}

The mean velocity $(U,V)$ fields are presented in figure \ref{fig:Mean2} for the $m=0$ cases and figure \ref{fig:Mean3} for the $m=1$ case. For the $\textit{NPR} = 2.10$ case, the transverse velocity due to the shocks does not exceed 3\% of the jet exit velocity $U_e$, while for the $\textit{NPR} = 3.4$ case the transverse velocity is in excess of 20\% of the jet exit velocity.

\subsection{Decomposition of Experimental Database}
The turbulent wavepackets that comprise the downstream-travelling component of aeroacoustic resonance typically only represent a small percentage of the total turbulent kinetic energy in a flow \citep{jordan2013wave,jaunet2017two,Towne2018spectral,schmidt2018spectral}. Eduction of the signatures of these wavepackets from experimental data thus requires some form of modal decomposition, such as those reviewed in \cite{taira2017modal}. 

Proper orthogonal decomposition (POD) is arguably the most widely-used decomposition method in fluid mechanics broadly \citep{berkooz1993proper,sirovich1987turbulence}, and flow resonance in particular \citep{MorenoPOD,DEMPOF2014,edgington2015multimodal,weightman2017explanation}. The highly-periodic nature of resonant flows makes them particularly amenable to POD, as it is typically possible to reconstruct the entire resonance cycle from only the leading POD modes \citep{KilianJFM}; a travelling wave structure will be defined by a pair of POD modes, with a $90^\circ$ phase offset between them \citep{deane1991low,noack2003hierarchy}. In the present database the velocity data is not time resolved, and thus decomposition such as spectral POD \citep{Towne2018spectral} cannot be implemented.

The ability of POD to isolate the structures associated with the resonant process enables a triple decomposition on the basis that the velocity may be represented as the sum of a mean ($U$), a coherent $u^{c}$ and a stochastic $u''$ component after \cite{HussainandReynolds1970},
\begin{equation}
\label{eq:tripleD}
\textbf{u}(\textbf{x},t) = \textbf{U}(\textbf{x})+\textbf{u}^{c}(\textbf{x},t)+\textbf{u}''(\textbf{x},t).
\end{equation}

To educe the coherent component via POD, an autocovariance matrix ($\textbf{R}$) is constructed from the velocity snapshots ($\textbf{V}$) such that $\textbf{R}=\textbf{V}^T\textbf{V}$. The solution of the eigenvalue problem $\textbf{R}\textbf{v}=\lambda \textbf{v}$ yields the eigenvalues $\lambda$ and eigenvectors $\textbf{v}$ from which the spatial POD modes ($\vect{\phi}$) are constructed as
\begin{equation}
\label{eq:spatialPOD}
\vect{\phi}_n(x,y)=\frac{\textbf{V}\textbf{v}_n(t)}{||\textbf{Vv}_n(t)||},
\end{equation}
and the coefficients at each time $t$ for each mode $n$ can be expressed as
\begin{equation}
\textbf{a}_n(t)=\textbf{v}_n(t)||\textbf{Vv}_n(t)||.
\end{equation}

Assuming that the leading pair of POD modes will identify fluctuations occurring at the screech frequency $\omega_s$,  we define \citep{jaunet2016pod}: $\textbf{a}=a_1 - ia_2=\hat{a}e^{-i \omega_s t}$ and $\vect{\psi} = \vect{\phi}_1 + i \vect{\phi}_2$. To ensure that the two leading modes are indeed the modal pair representing screech, the decomposition is conducted only on the transverse component of velocity \citep{weightman2018signatures}. On this basis the coherent fluctuations can be represented as
\begin{equation}
\textbf{q}^c(x,y,t) = \hat{a}e^{-i \omega_s t} \vect{\psi}(x,y).
\end{equation}

To identify wavelike structures in the flow, it is advantageous to consider a further decomposition in the streamwise direction,
\begin{equation}
\textbf{q}^c(x,y,t) = \hat{a}e^{-i \omega_s t}\sum_{k} \hat{\textbf{q}}^c_k(y)e^{ikx}.
\end{equation}
Here, the temporal Fourier coefficients have been constructed directly from the complex POD mode pair $\vect{\psi}$, such that
\begin{equation}
\label{eq:FD}
\hat{\textbf{q}}^c_k(y) = \sum_{x}\vect{\psi}(x,y)e^{ikx}.
\end{equation}

\subsection{Global stability analysis}
We conduct a global linear stability analysis to explore the characteristics and behavior of the waves involved in the screech process.  Applying a Reynolds decomposition 
\begin{equation}
\label{Eq:reynolds_decomp}
\mathbf{q}(x,r,\theta,t) = \bar{\mathbf{q}}(x,r) + \mathbf{q}^{\prime}(x,r,\theta,t)
\end{equation}
to the compressible Navier-Stokes equation and neglecting nonlinear terms yields the linearized Navier-Stokes equation,
\begin{equation}
\label{Eq:linearNS}
\frac{\partial \mathbf{q}^{\prime}}{\partial t} - \mathbf{A}(\bar{\mathbf{q}}) \, \mathbf{q}^{\prime} = \mathbf{0}.
\end{equation}
Here, $\mathbf{q}(x,r,\theta,t)$ is a state vector containing velocities and thermodynamic variables; for the round jets considered in this paper we use cylindrical coordinates and velocities and choose density and pressure as the thermodynamic variables.  Applying the normal mode ansatz
\begin{equation}
\label{Eq:normal_modes_global}
\mathbf{q}(x,r,\theta,t) = \hat{\mathbf{q}}(x,r) \exp\left(i m \theta - i \omega t  \right),
\end{equation}
to~(\ref{Eq:linearNS}) yields the eigenvalue problem
\begin{equation}
\label{Eq:LNS_global}
\left(- i \omega \mathbf{I}  - \mathbf{A}_{m} \right) \hat{\mathbf{q}} = \mathbf{0},
\end{equation}
where $\mathbf{A}_{m}$ is obtained by replacing all azimuthal derivatives in $\mathbf{A}$ with $im$.  The azimuthal wavenumber $m$ is an integer due to the periodicity of the mean jet.  The global modes of the jet correspond to $\omega$, $\hat{\mathbf{q}}$ pairs that satisfy~(\ref{Eq:LNS_global}) for a given choice of $m$, i.e., the eigenvalues and eigenvectors of $\mathbf{A}_{m}$.

Significant uncertainty remains regarding the effect of shocks and shock-like discontinuities on the validity of a linear stability analysis; we seek to minimize any such effects in two ways. Firstly, we consider only the $\textit{NPR} =2.10$ case, where the shocks are weak, and the transverse velocities are minimal. Much of the compression in a jet operating at this condition is achieved in a continuous fashion through near-isentropic compression waves. Additionally, the inability of tracer particles to faithfully reproduce step changes in velocity is actually of benefit here; the discontinuities around shocks are inherently smoothed by our measurement technique. 

The mean flow $\bar{\mathbf{q}}(x,r)$ is obtained from the experimental measurements.  To provide a sufficiently large domain for the linear analysis, the experimental domain is extrapolated to a field covering $-1 \leq x/D \leq 19$ and $0 \leq r/D \leq 4$. Density information is not directly available from PIV, however by assuming that entropy generation by the weak shocks is negligible, an assumption of constant stagnation density along streamlines allows for a calculation of density from the velocity data. This is checked using a tomographic Background-Oriented-Schlieren measurement, with the axisymmetric field extracted from the path-integrated data using an Abel inversion \citep{tan2015measurement}; errors due to the isentropic assumption appear to be less than 5\%. 

\begin{figure}
\centering
\includegraphics[width=0.9\textwidth]{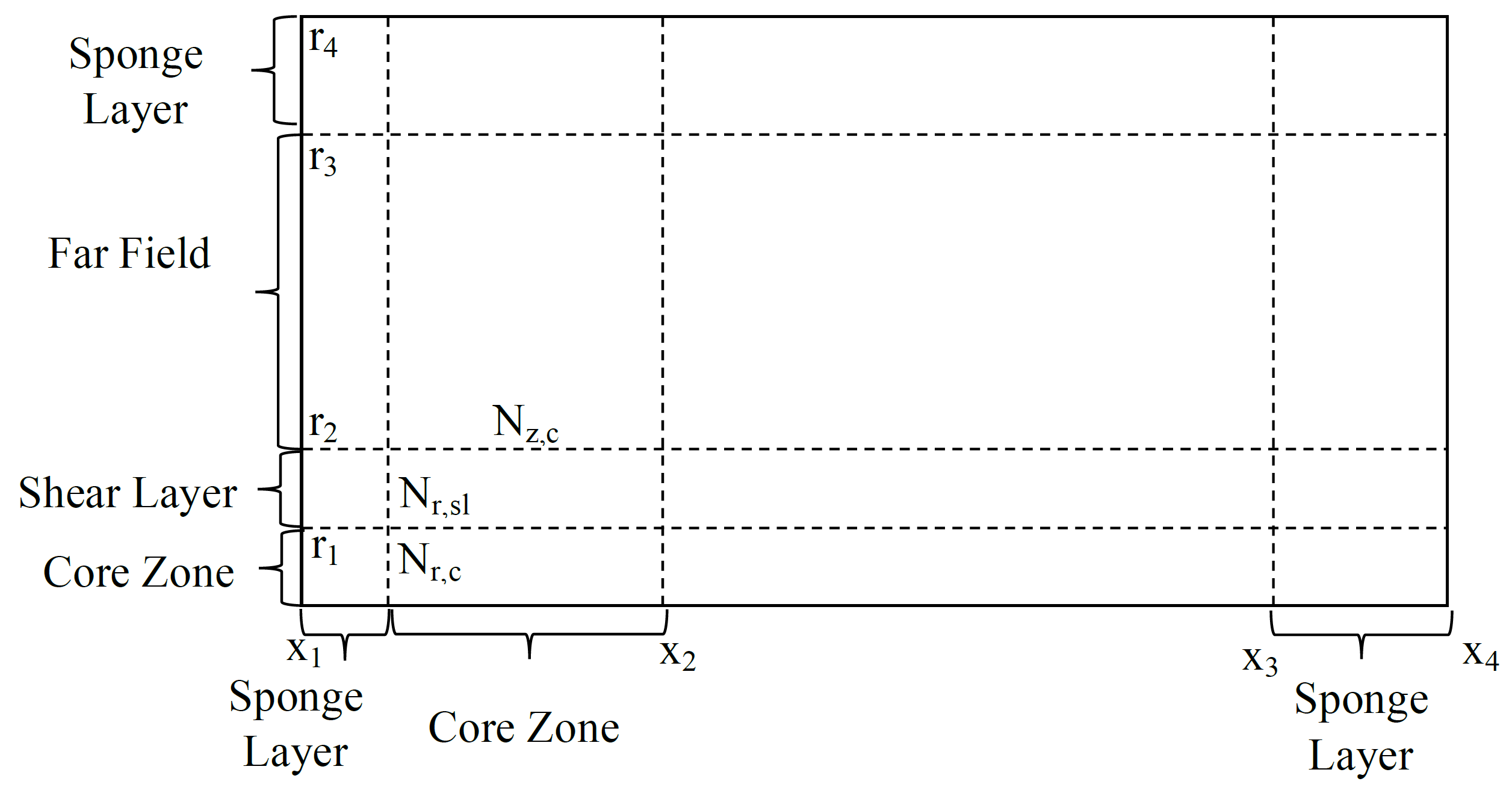}
\caption{Calculation domain and mesh structure for the global analysis.}
\label{fig:GS_Dom}
\end{figure}

The linearized equations are discretized using fourth-order finite differences with summation-by-parts closure at the boundaries.  To ensure adequate resolution in the shear layer of the jet and to apply non-reflecting boundary conditions, the computational domain is divided into several zones as shown in figure \ref{fig:GS_Dom}. The core domain covers $0\leq x/D \leq 8$ and $0 \leq r/D \leq 0.3$, the shear layer zone covers $0.3 \leq r/D \leq 0.7$, and the far-field extends to $r/D = 3$. Sponge layers are placed upstream ($x/D \leq 0$) and downstream ($x/D \geq 15$) of the jet, as well as outside the far-field region. There are $N_r=300$ points in the radial direction: 100 in the core zone, 100 in the shear layer, and the rest shared between the far-field and the sponge layer.  The streamwise direction is discretized using $N_x=750$ points with a mapping was used to ensure that $60\%$ of these points are placed in the core region.  Addition details of the global stability code can be found in \cite{schmidt2017wavepackets}.

The analysis was performed at a lower Reynolds number compared to the experiments ($Re=\rho_j U_j D/\mu=10^3$, where $\rho_j$ and $U_j$ are the ideally expanded density and velocity, and $\mu$ is the dynamic viscosity).  This choice was motivated by recent work demonstrating that the use of an eddy viscosity or turbulent Reynolds number to account for the impact of Reynolds stresses improves the agreement between linear analyses and experimental and simulation data for turbulent jets \cite{pickering2020optimal}.  The lower Reynolds number has the added advantage of reducing the cost of solving the eigenvalue problem.

\begin{figure}
\centering
\includegraphics[width=\textwidth]{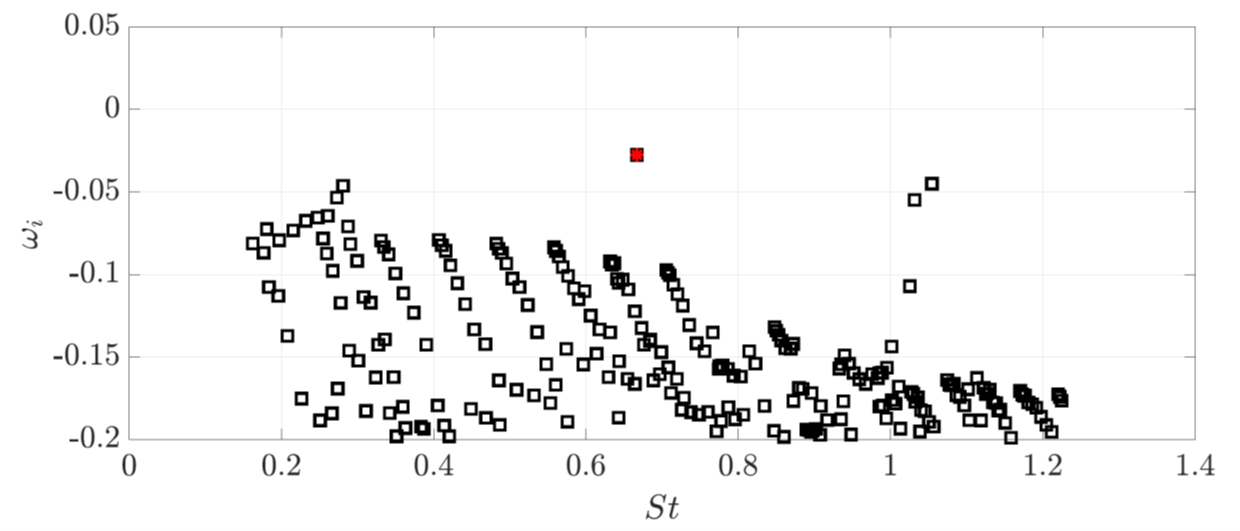}
\caption{Eigenspectrum from global stability analysis for the $\textit{NPR} = 2.10$ jet. Red star indicates most unstable mode at $St = 0.667$.}
\label{fig:G_Eig}
\end{figure}

The eigenvalue spectrum obtained from the global eigenvalue problem is presented in figure \ref{fig:G_Eig}. The spectrum is composed of continuous branches of modes plus a few discrete modes, which are distinct from the main branches and arise at particular frequencies. The least-stable mode is associated with a frequency of $St = 0.667$, very close to the experimentally observed screech tone at frequency $St = 0.67$ for this jet. The remainder of the analysis focuses on this most-unstable mode.

\subsection{Local stability analysis}
We also conduct a local stability analysis of the same mean flow to aid in identifying the waves observed in the experimental POD and global stability modes.  In this case, the appropriate normal mode ansatz is
\begin{equation}
\label{Eq:normal_modes}
\mathbf{q}(x,r,\theta,t) = \hat{\mathbf{q}}(r) \exp\left( i k x + i m \theta - i \omega t  \right),
\end{equation}
where $k$ is the streamwise wavenumber for the mode.  Setting the mean flow within the operator $\mathbf{A}$ to its local value, $\bar{\mathbf{q}}(x=x_0,r)$, at a particular streamwise position $x_0$ and applying the ansatz~(\ref{Eq:normal_modes}) to~(\ref{Eq:linearNS}) leads to an eigenvalue problem of the form
\begin{equation}
\label{Eq:LNS_local_eig}
\left( - i \omega \mathbf{I} + i k \mathbf{A}_{x} + \mathbf{A}_{m,0} \right) \hat{\mathbf{q}} = 0.
\end{equation}
The operator $\mathbf{A}_{x}$ is the portion of $\mathbf{A}$ associated with $x$-derivatives while $\mathbf{A}_{m,0}$ contains all remaining terms.  For simpicity, second steamwise derivatives have been neglected in order to obtain an eigevalue problem that is linear in $\alpha$.  Both $\mathbf{A}_{x}$ and $\mathbf{A}_{m,0}$ depend on the position $x_0$ at which the mean flow has been frozen.  For a specified $m$, the local eigenmodes of the jet are given by pairs $(k,\omega)$ and vectors $\hat{\mathbf{q}}$ satisfying~(\ref{Eq:LNS_local_eig}).  The spatial stability analysis used in this paper is carried out by choosing real values for the frequency $\omega$ and solving the eigenvalue problem for the (potentially complex-valued) wavenumber $k$. Details on the operators and numerics used for the eigenvalue problem are reported elsewhere \citep{towne2016advancements}.

\begin{figure}
\centering
\includegraphics[width=0.7\textwidth]{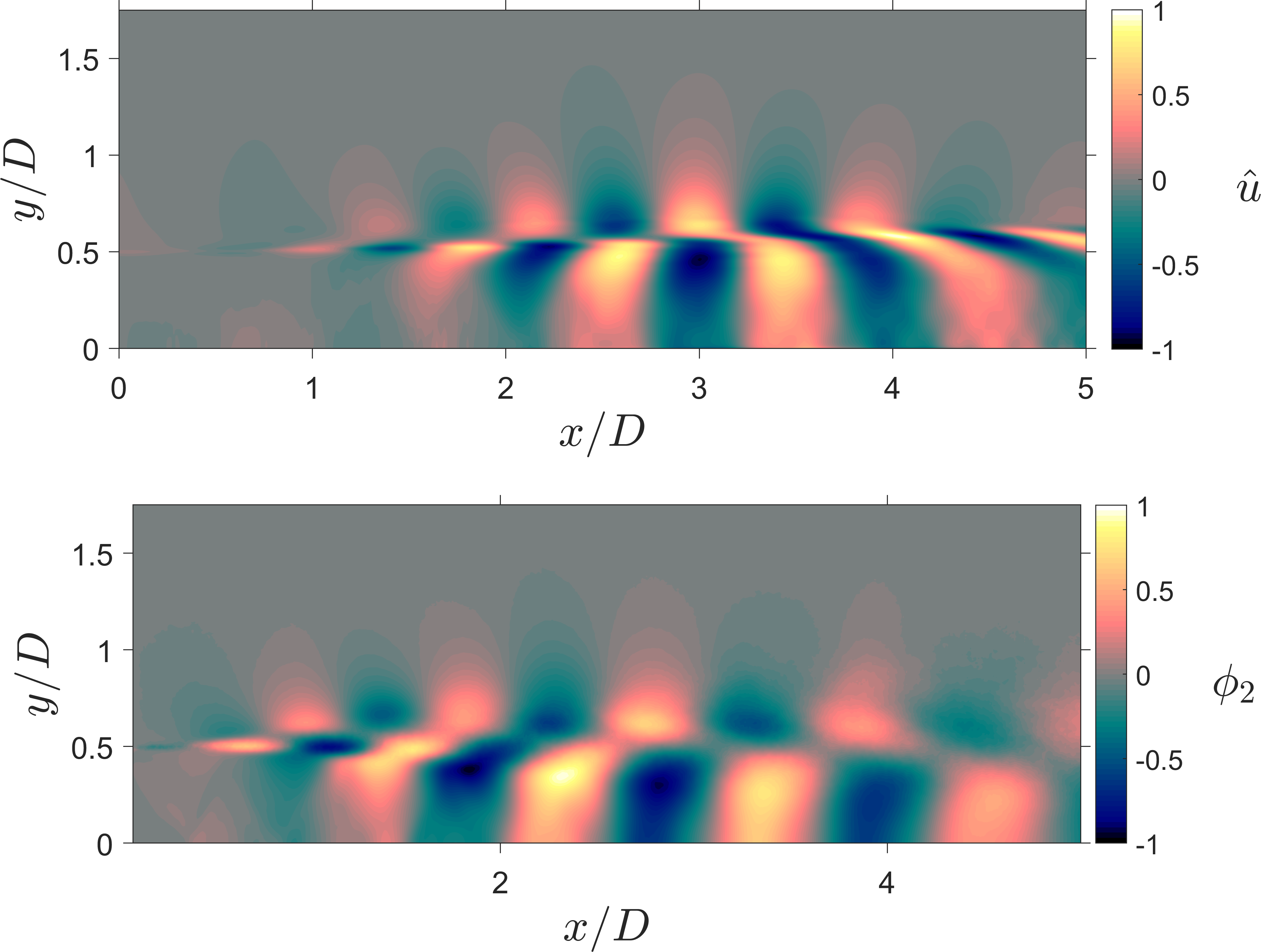}
\caption{Streamwise velocity (top) associated with the global mode identified at $St = 0.667$, presented with an experimentally-determined POD streamwise velocity mode (bottom). All modes are normalized against their maximum value. Results are for the $\textit{NPR} = 2.10$ jet.}
\label{fig:G_POD}
\end{figure}

\begin{figure}
\centering
\includegraphics[width=1\textwidth]{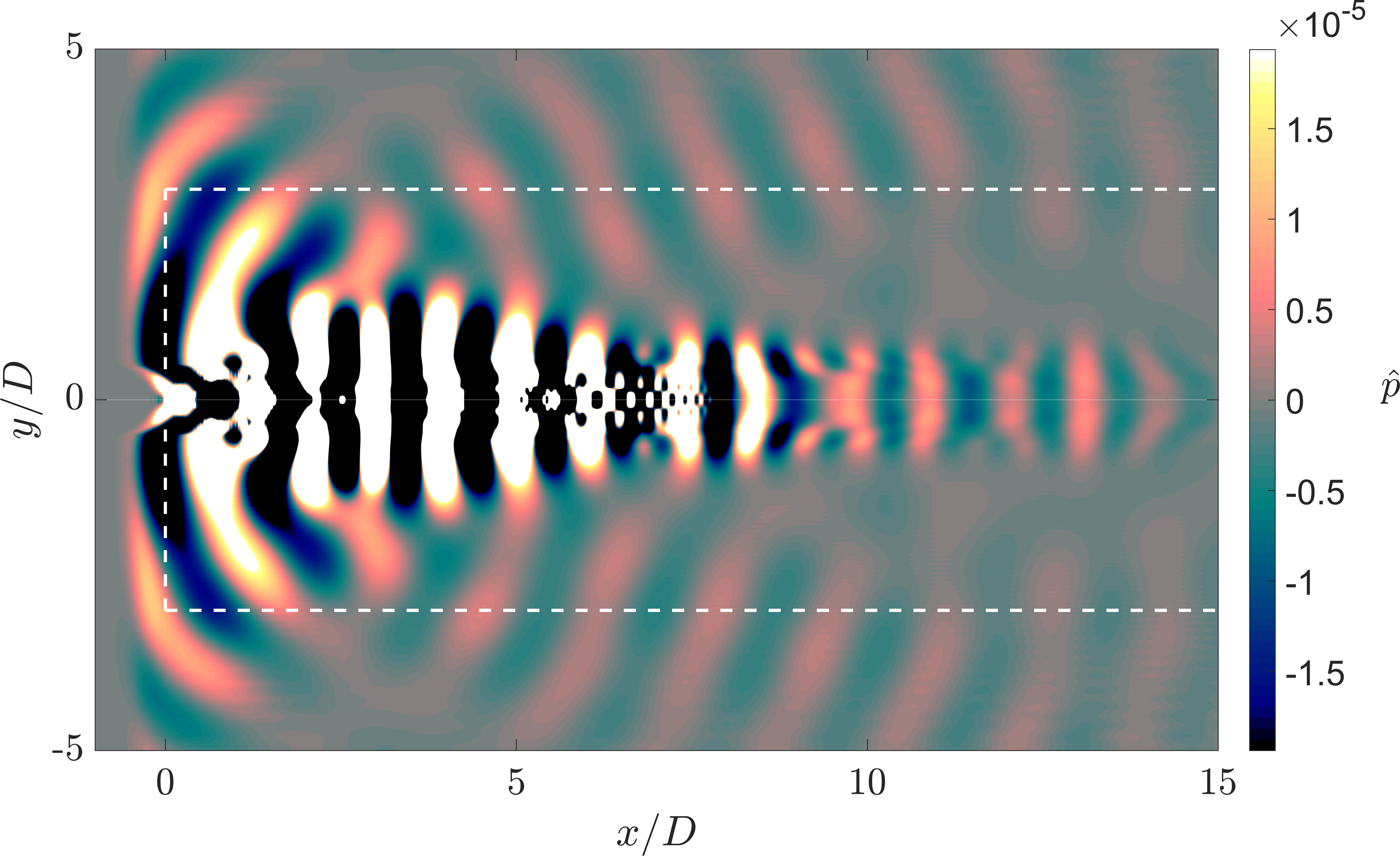}
\caption{Pressure field associated with the least-stable global mode. The dashed white lines indicate the beginning of the sponge zones; outside these bounds the contours of pressure are only included as a qualitative visualization.}
\label{fig:G_PRS}
\end{figure}

\section{Results}
\subsection{Wave generation in screeching jets}
In figure \ref{fig:G_POD} we compare the spatial structure of the most unstable mode in the global analysis to one of the leading POD modes extracted from the experimental data. In this work the radial co-ordinate $r$ from the stability analysis is represented as the transverse co-ordinate $y$ for consistency with the PIV measurements. A remarkable similarity between the experimental POD mode and the global mode is apparent. While the streamwise velocities are not identical, given that the global mode was calculated from a linearization of an extrapolated base flow, the agreement is surprisingly good. In figure \ref{fig:G_PRS} the pressure field associated with the same global mode is presented. While spatially resolved pressure data are not available for the measurement, the pressure field captures many of the recognized features of a screeching supersonic jet, including the strong upstream-propagating acoustic waves, and the downstream-propagating Mach wave radiation.

\begin{figure}
\centering
\includegraphics[width=1\textwidth]{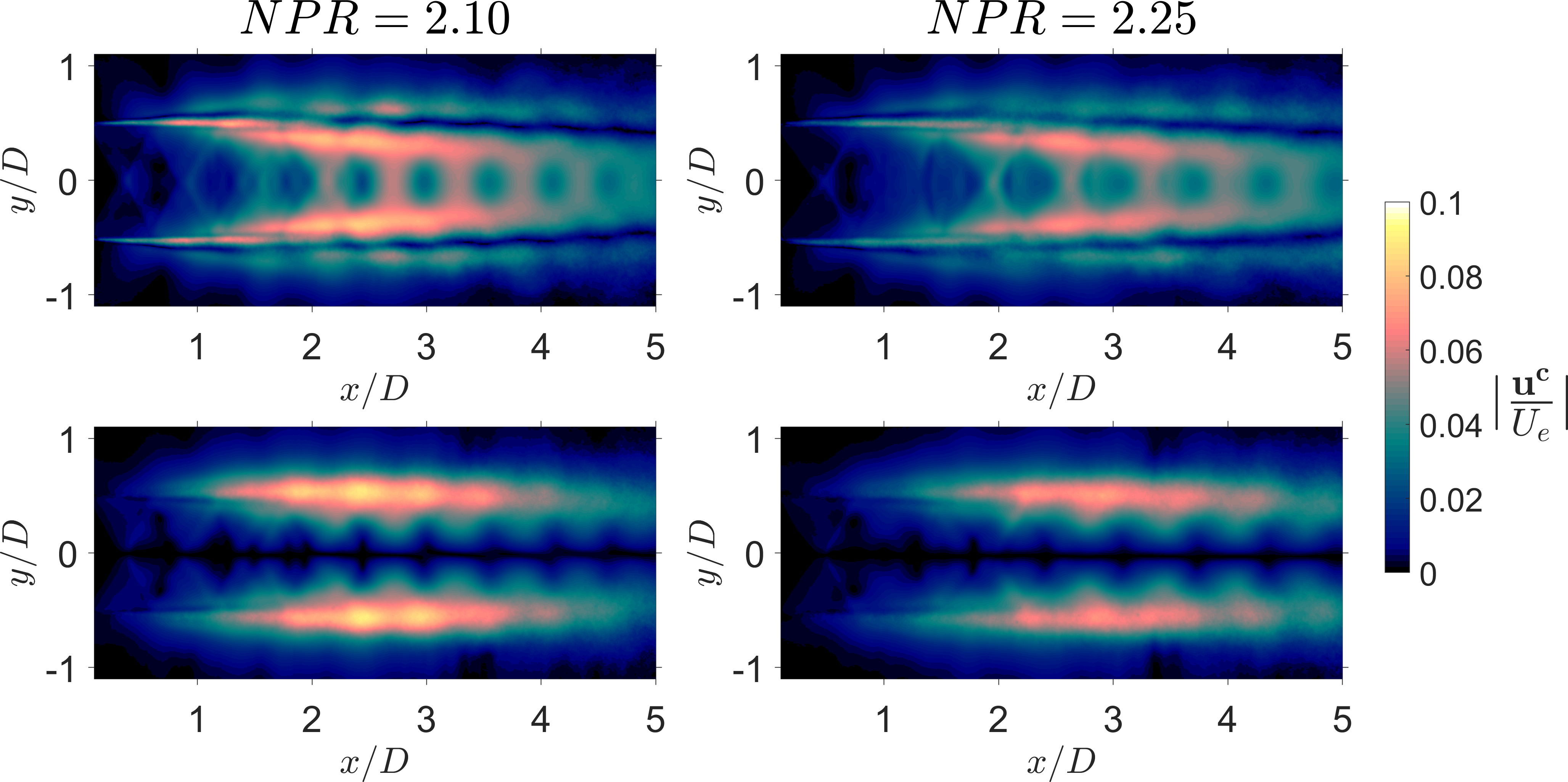}
\caption{Magnitude of coherent fluctuations for the two jets where the screech mode is an m = 0 azimuthal mode. Upper) Axial velocity fluctuations. Lower) Transverse velocity fluctuations.}
\label{fig:Coh1}
\end{figure}

\begin{figure}
\centering
\includegraphics[width=1\textwidth]{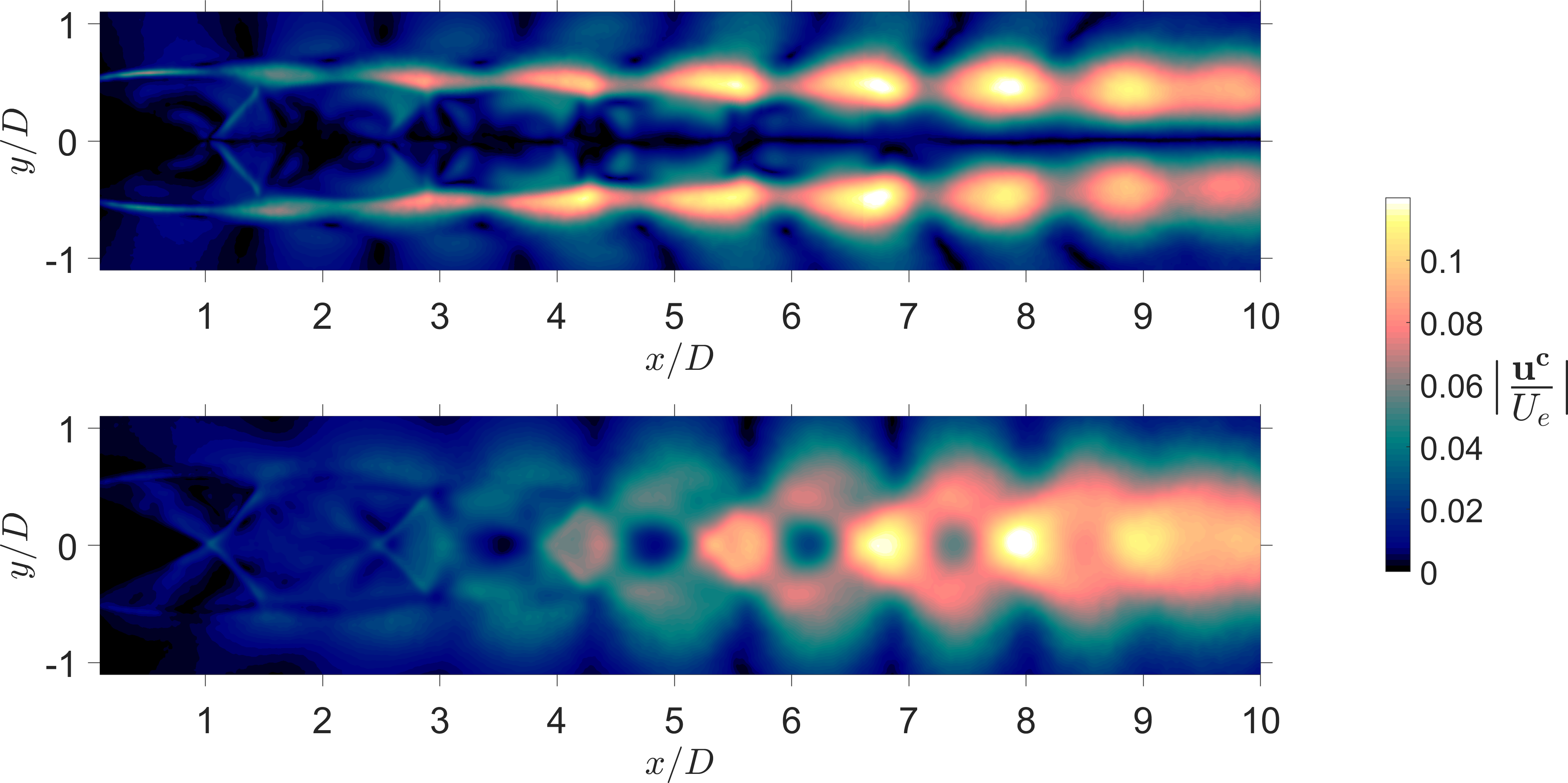}
\caption{Magnitude of coherent fluctuations for jet where the screech mode is an m = 1 azimuthal mode. Upper) Axial velocity fluctuations. Lower) Transverse velocity fluctuations.}
\label{fig:Coh2}
\end{figure}

\begin{figure}
\centering
\includegraphics[width=1\textwidth]{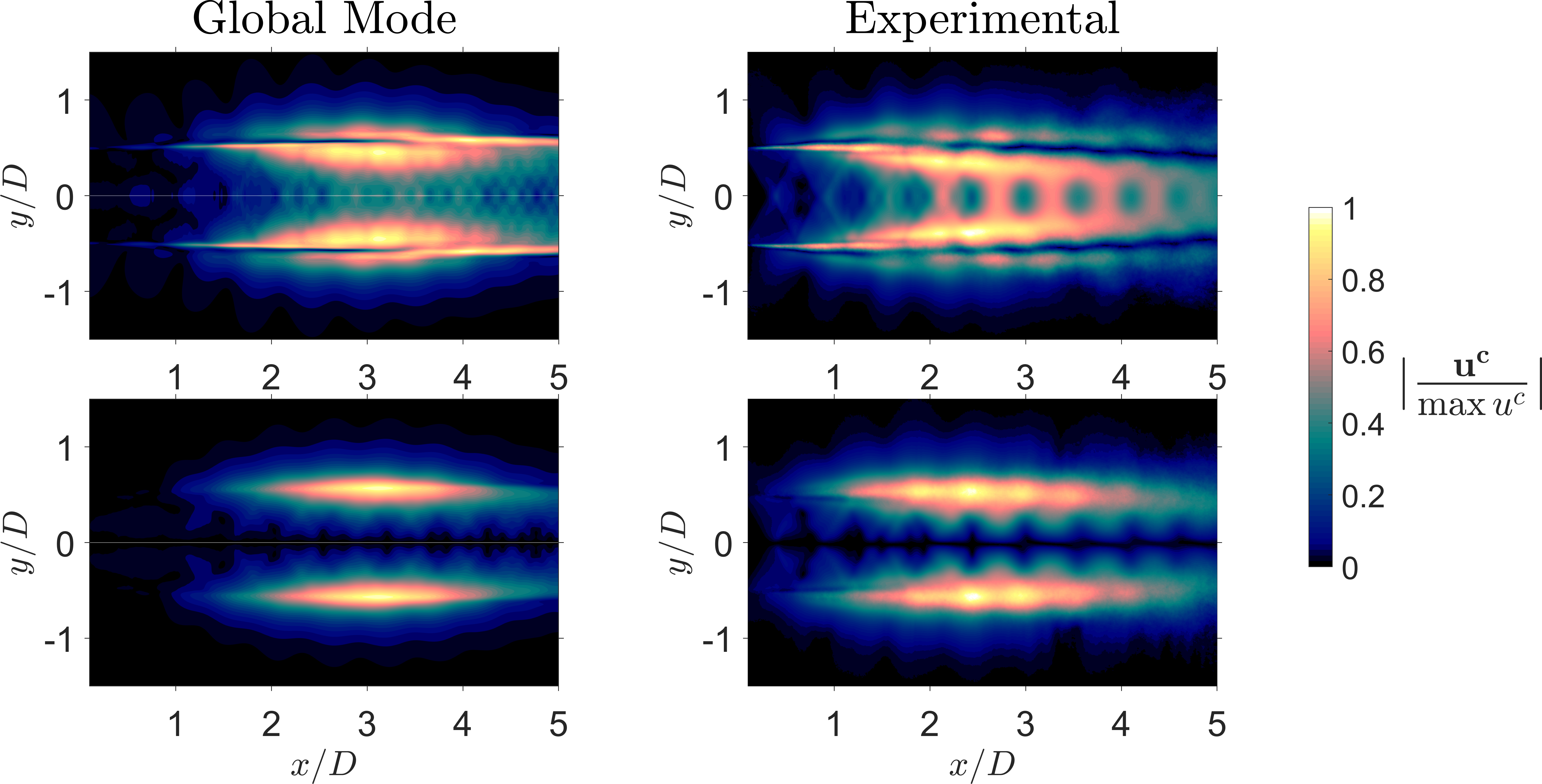}
\caption{Magnitude of fluctuations from both experiment and global analysis. Upper) Axial velocity fluctuations. Lower) Transverse velocity fluctuations. All results for $\textit{NPR} = 2.10$ jet.}
\label{fig:G_A}
\end{figure}

The normalized coherent fluctuations from the experimental data $|\vect{\psi}|$ for the $m=0$ jets is presented in figure \ref{fig:Coh1}, and for the $m=1$ jet in figure \ref{fig:Coh2}. The coherent fluctuations are quite similar for the two $m=0$ modes, with spatial modulation apparent at both the edges of the shear layer and along the jet centreline. For the $m=1$ jet, modulation of the axial velocity component is evident both along the lipline and outside the shear layer, while for the transverse velocity component the strongest modulation occurs along the centreline. A comparison of the fluctuations determined from the most unstable global mode with those associated with the leading POD mode pair is provided in figure \ref{fig:G_A}. Qualitatively, the two results are very similar. The streamwise velocity experiences modulation in the jet core and outside the shear layer for both the global analysis and the experimental data. The transverse velocity is even more closely matched. There are however also some key differences: the magnitude of fluctuations in the jet core is higher for the experimental data, and their modulation stronger.

\begin{figure}
\centering
\includegraphics[width=1\textwidth]{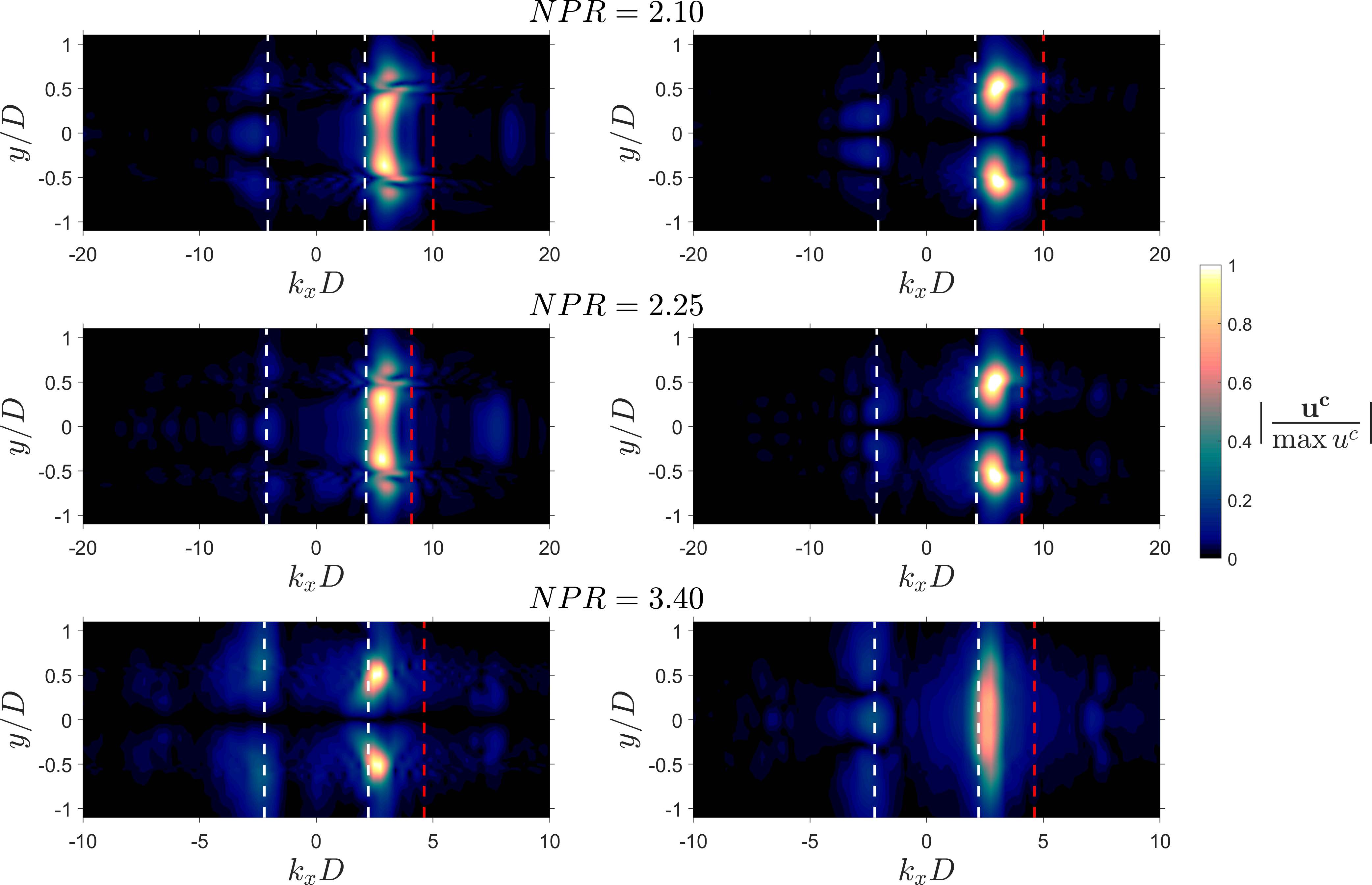}
\caption{Wavenumber spectra for all three jets. Left) Axial velocity fluctuations. Right) Transverse velocity fluctuations. Note the different x-axis scale for $\textit{NPR} = 3.4$.}
\label{fig:WNS}
\end{figure}

\begin{figure}
\centering
\includegraphics[width=1\textwidth]{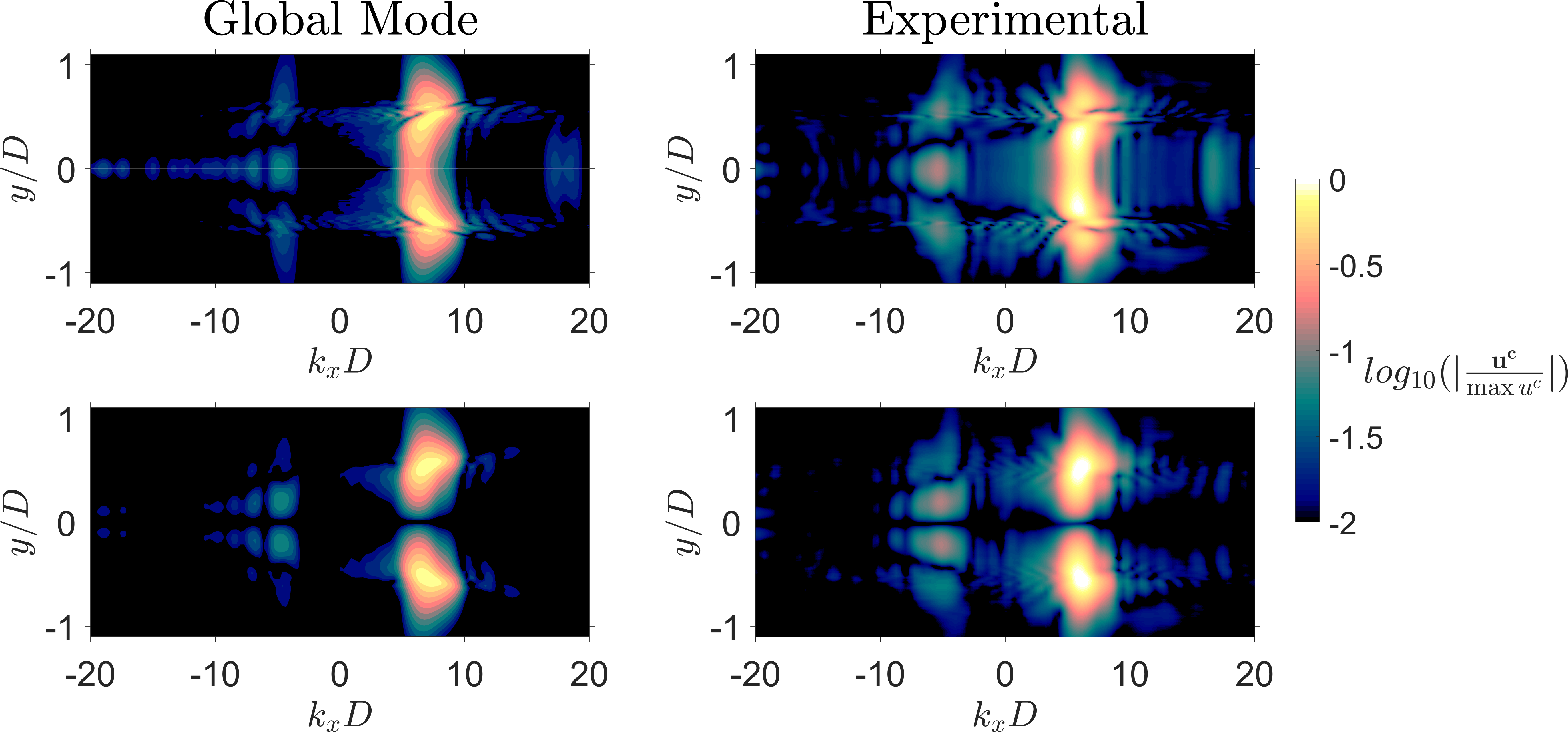}
\caption{Wavenumber spectra from both global mode and experiment. Upper) Axial velocity fluctuations. Lower) Transverse velocity fluctuations. All results are for the $\textit{NPR} = 2.10$ jet. Note that a logarithmic contour scale is used in this image.}
\label{fig:G_WNS}
\end{figure}

While the fluctuations presented in the preceding figures are all associated with fluctuations at a given frequency, these fluctuations can be associated with a broad range of wavenumbers. Consequently, the wavenumber spectra presented in figure \ref{fig:WNS} are produced by taking the amplitude of (\ref{eq:FD}). Phase velocity is defined as $u_p=\omega / k_x$; as the POD produces modes correlated to the screech phenomenon, here $\omega$ is fixed at the screech frequency: $\omega = \omega_s$. Thus the sign of $k_x$ determines the sign of the phase velocity; here positive values of $k_x$ are associated with a phase velocity in the downstream direction. In this work, we use the sign of the phase velocity as a proxy for the sign of the group velocity. The group velocity determines the direction of energy propagation; waves with negative and positive group velocity may be considered upstream- or downstream-travelling respectively. All of the waves in question have phase and group velocities of the same sign in supersonic jets \citep{TowneetalJFM2017}, justifying the assumption made in this analysis. The dashed vertical white lines in figure \ref{fig:WNS} indicate wavenumbers associated with the ambient speed of sound in the upstream and downstream directions, while the dashed vertical red line indicates the wavenumber associated with the average spacing of the shocks in the flow. All three jets have the majority of the energy concentrated at a wavenumber associated with a phase velocity of $u_p \approx 0.7U_j$, with radial structures typical of the classical KH wavepacket, hereafter referred to as the $k^+_\textit{kh}$ wave. All three jets also have a component with an upstream phase velocity approximately equal to the speed of sound; this is the signature of the acoustic mode previously documented in screeching jets \citep{gojon2018oscillation,edgington2018upstream}, hereafter the $k^-_\textit{th}$ wave. There is evidence of a third wave at much higher positive wavenumber, observed for all jets, though with different radial structure for the $m=1$ case compared to the $m=0$ screech modes; this wave will be referred to as the $k^+_{t}$ wave. The spectrum of the global mode analysis is cleaner than that of the experimental data as shown in figure \ref{fig:G_WNS}, but the same three structures visible in the experimental data are likewise visible: an upstream-travelling acoustic mode, a downstream-travelling KH wave, and the high-wavenumber mode. The relative amplitudes (normalized against the $k^+_\textit{kh}$ wave) for both the $k^-_\textit{th}$ and the $k^+_{t}$ mode are significantly stronger in the experimental data, but still clearly visible in the global mode. In the global analysis, the wavenumbers for all three waves are approximately $\Delta k_{x} D = 0.5$ higher than the corresponding wave in the experimental data.

We return now to the model of \cite{tam1982shock}, where it was suggested that the interaction of the KH wavepacket with the stationary shock structures should produce two additional travelling waves in the jet. Due to variations in the mean flow, the fluctuations associated with the KH wavepacket are spread across a small range of wavenumbers. Likewise, the shock spacing within the jet varies slightly as a function of axial position. Estimates of the wavenumbers associated with both the KH wavepacket and the shock cells are presented in table \ref{tab:WN}, with a rough estimate of the variation in wavenumber included. We then evaluate (\ref{eq:TT2}) to produce the two expected wavenumbers (sum and difference) resulting from the interaction between the KH wavepacket and the shock cells. These expected wavenumbers are presented in table \ref{tab:WND}, along with wavenumbers extracted from the the data presented in figure \ref{fig:WNS}. The observed wavenumbers for all the experimental cases and the global analysis fall within the range of expected wavenumbers produced by (\ref{eq:TT2}). Thus the model of \cite{tam1982shock} provides an explanation for the three wave structures observed in the decomposed experimental data and in the global analysis: one structure is the KH wavepacket, while the other two are waves produced by the interaction between the KH wavepacket and the quasi-stationary shock cells. While the model of \cite{tam1982shock} predicts the peak wavenumbers of these waves, and provides an explanation for their mechanism of generation, it makes no statement regarding the character of these waves. A cursory examination of figure \ref{fig:WNS} reveals that each of the three modes has a distinct radial structure. In the following section, we consider the nature of these three waves.

\begin{table}
\centering
\caption{Wavenumbers of wave structures}
\label{tab:WN}       
\begin{tabular}{cccc}
\hline\noalign{\smallskip}
\textit{NPR}  &  $k_\textit{kh} D $  &  \quad \quad \quad &   $k_s D$   \\
\noalign{\smallskip}\hline\noalign{\smallskip}
2.10 (Exp) & $5.8 \pm 0.5$  & & $11 \pm 0.5$ \\
2.10 (LSA) & $6.2 \pm 0.5$ & & $11 \pm 0.5$ \\
2.25       & $5.5 \pm 0.5$ & & $9 \pm 0.5$ \\
3.4        & $2.6 \pm 0.3$ & & $4.6 \pm 0.2$\\
\noalign{\smallskip}\hline
\end{tabular}
\end{table}

\begin{table}
\centering
\caption{Wavenumber differences}
\label{tab:WND}       
\begin{tabular}{cccccc}
\hline\noalign{\smallskip}
\textit{NPR}  & $k_\textit{kh}D-k_sD$   &   $k_x D (k^-_\textit{th})$ & \quad \quad \quad& $k_\textit{kh}D+k_sD$ & $ k_x D (k^+_{t})$  \\
\noalign{\smallskip}\hline\noalign{\smallskip}
2.10 (Exp) & $-5.2$  &  $-5.5$ &\quad & $16.8$ & $16.7$  \\
2.10 (LSA) & $-4.8$  &  $-4.9$ &\quad & $17.2$ & $17.8$ \\
2.25       & $-4.0$  &  $-3.5$ &\quad & $14.5$ & $15.3$\\
3.4        & $-2.0$  &  $-2.3$ &\quad & $7.2$  & $7.2$\\
\noalign{\smallskip}\hline
\end{tabular}
\end{table}

\subsection{The nature of waves in screeching jets}
\begin{figure}
\centering
\includegraphics[width=1\textwidth]{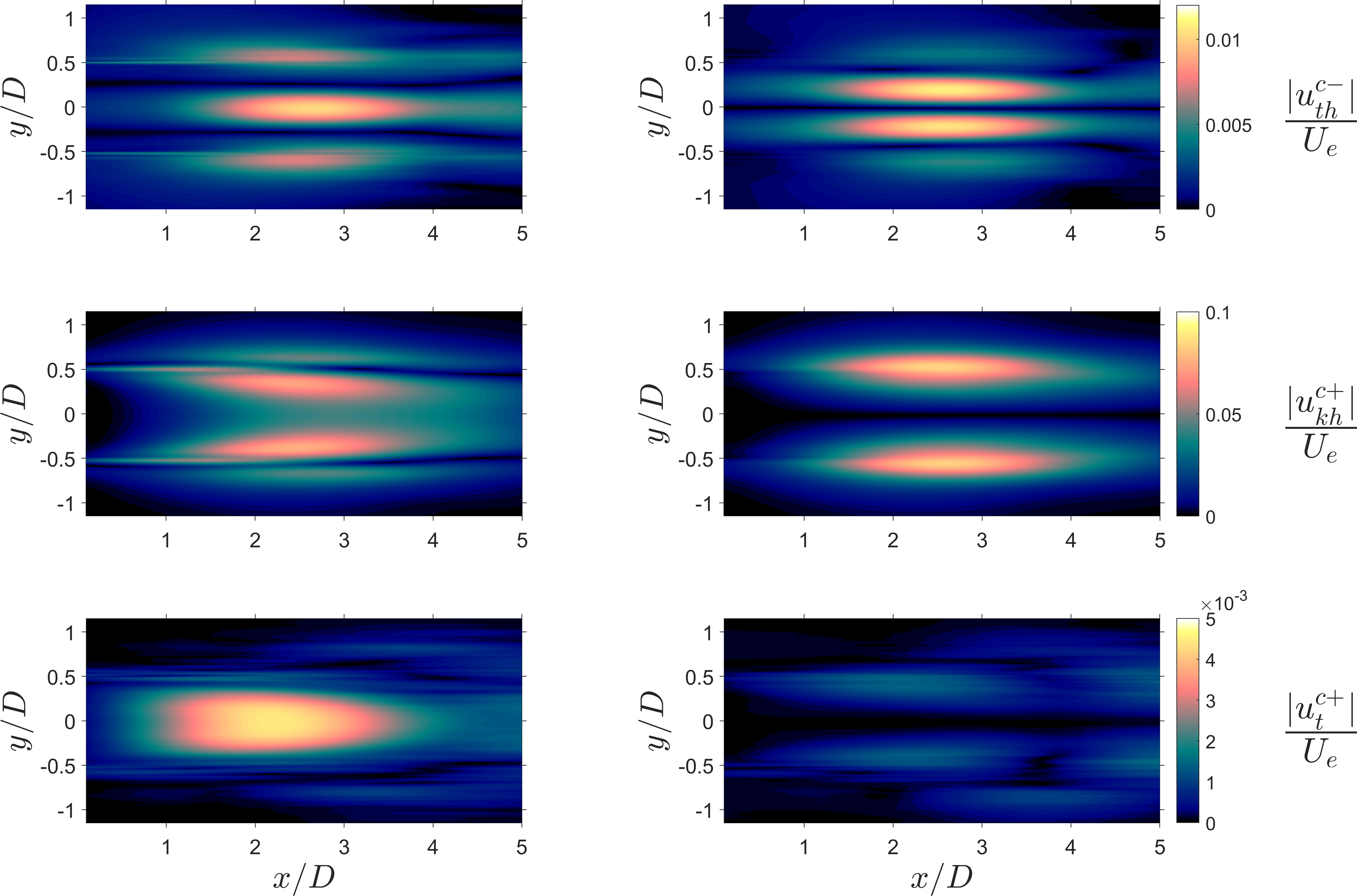}
\caption{Qualitative amplitude distributions associated with the three wavelike structures for the \textit{NPR} = 2.10 jet (experimental). Top) $k^-_\textit{th}$. Middle) $k^+_\textit{kh}$. Bottom) $k^+_{t}$. Left) Axial velocity fluctuations. Right) Transverse velocity fluctuations.}
\label{fig:NPR210wamp}
\end{figure}

\begin{figure}
\centering
\includegraphics[width=1\textwidth]{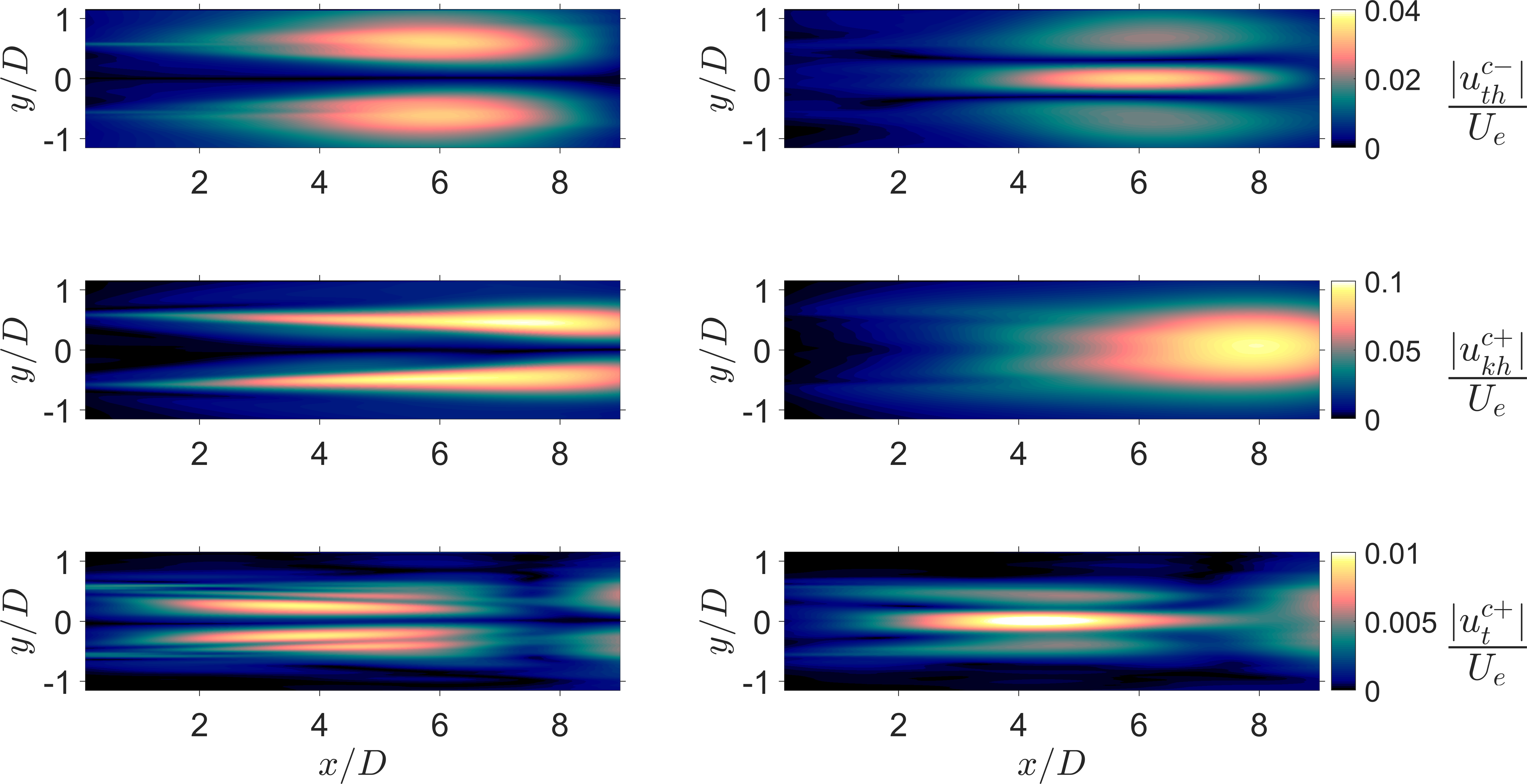}
\caption{Qualitative amplitude distributions associated with the three wavelike structures for the \textit{NPR} = 3.40 jet (experimental). Top) $k^-_\textit{th}$. Middle) $k^+_\textit{kh}$. Bottom) $k^+_{t}$. Left) Axial velocity fluctuations. Right) Transverse velocity fluctuations.}
\label{fig:NPR34wamp}
\end{figure}

To better characterize the three wavelike structures in the jet, we consider their spatial features: first their axial variation in amplitude and then their radial structure. The spatial amplitude variation associated with the waves is shown qualitatively via the application of a cosine-tapered bandpass filter in the wavenumber domain. The filter has a half-width of $1.9D$, a taper ratio of $0.5$, and is centred on the maximum amplitude for each wave. The results are sensitive to the size and type of the filter; these results should thus only be taken as indicative of peak location and general trends. The amplitude of the fluctuations resulting from this filtering is presented for $\textit{NPR} = 2.10$ in figure \ref{fig:NPR210wamp} and for $\textit{NPR} = 3.40$ in figure \ref{fig:NPR34wamp}. The spatial amplitude distribution of the $k^+_\textit{kh}$ wavepacket closely resembles the distribution in figures \ref{fig:Coh1} \& \ref{fig:Coh2}, but without the spatial modulation. This result is unsurprising; the majority of the downstream-travelling energy is associated with the KH wave, and modulation cannot be represented with a single wave. The $k^{-}_\textit{th}$ wave peaks at approximately the fourth shock cell for both jets (often observed as the peak sound source location in screeching jets \citep{mercier2017experimental}), and as previously discussed in \cite{edgington2018upstream} has support outside the shear layer and in the core of the jet. The $k^+_{t}$ downstream wave reaches maximum amplitude slightly upstream of the peak amplitude location for the upstream wave in the $\textit{NPR} = 2.10$ jet, but somewhat further upstream for the $\textit{NPR} = 3.40$ jet. The fluctuations associated with this wave are almost entirely bound within the core of the jet. At $\textit{NPR} = 2.10$ the amplitude of the axial fluctuations are significantly larger than the transverse; at $\textit{NPR} = 3.40$ both transverse and axial fluctuations are observed for the $k^+_{t}$ wave. 

\begin{figure}
\centering
\includegraphics[width=1\textwidth]{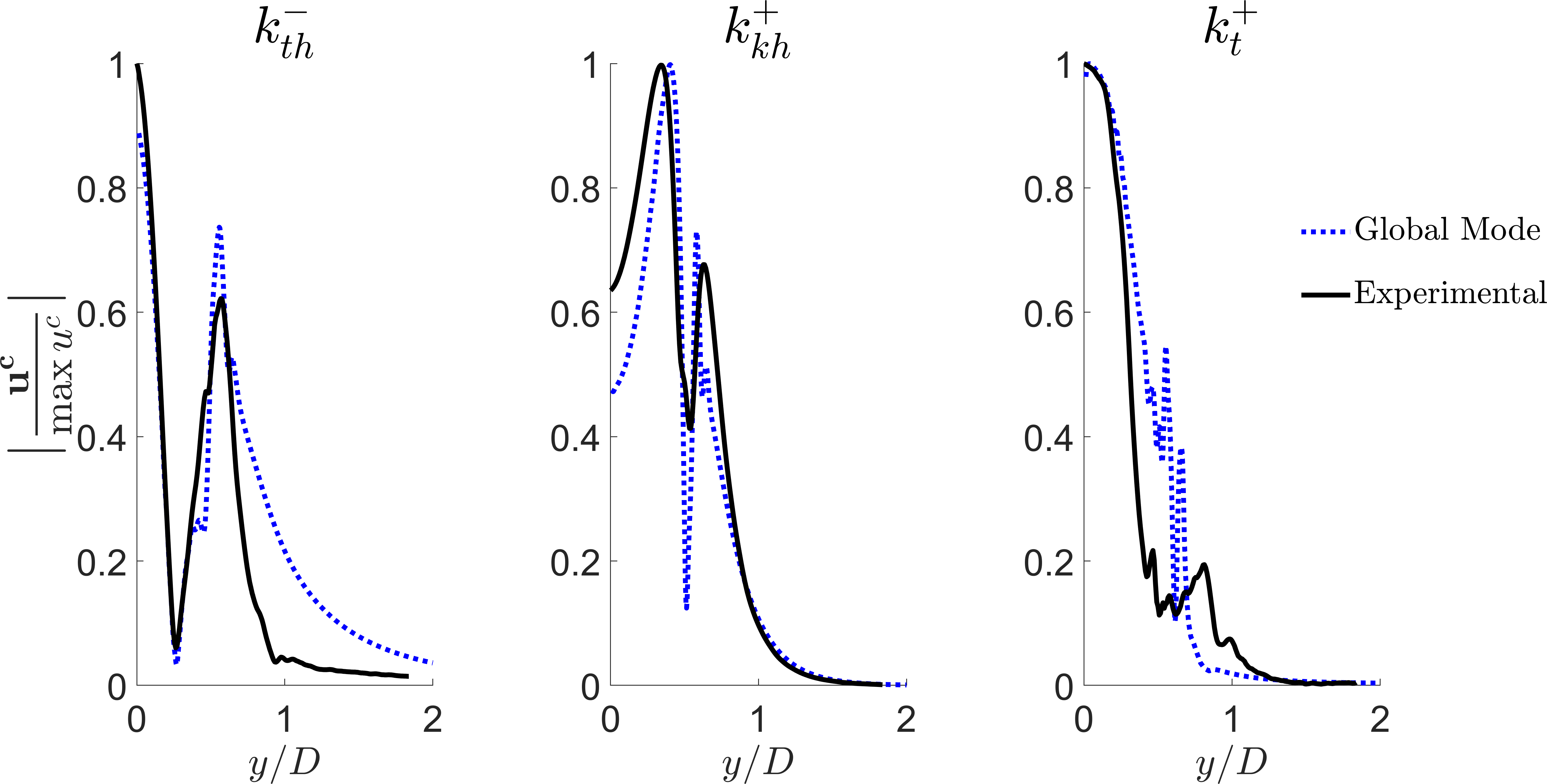}
\caption{Radial profiles of streamwise velocity for the three modes discussed in this paper. All modes are normalized to their own maximum value.  All results are for the $\textit{NPR} = 2.10$ jet.}
\label{fig:rad1}
\end{figure}

A comparison of the radial profiles of streamwise velocity for the three identified structures is presented in figure \ref{fig:rad1}, educed from both experiment and global analysis for the $\textit{NPR} = 2.10$ jet. All curves have been lightly smoothed with a moving average filter to reduce noise; no major features have been removed. The radial structure of the $k^-_\textit{th}$ wave exhibits a close match between theory and experiment, though the radial decay outside the jet is slower for the global mode. This mismatch in the acoustic field may be a consequence of the PIV measurement being unable to capture weak acoustic perturbations further from the jet. For the $k^+_\textit{kh}$ wave, the radial structure for both experiment and LSA exhibits the characteristic double peak of a KH wavepacket, with minima at the same radial location, but with a closer spacing between the peaks for the global mode. The radial structure of the $k^+_{t}$ mode is very similar for the two analyses, though there is a small peak at $r/D \approx 0.5$ for the global mode that is larger than that observed in the experiment. As noted earlier, the wavenumber for each of the modes identified in the global analysis is about $\Delta k_{x} / D = 0.5$ higher than the corresponding wave in the experimental data. Since the global analysis is performed on the experimental mean flow, the spacing of the shock cells is the same for both cases. Given this fact, and the slightly higher wavenumber for the KH wave in the global analysis, the increase in wavenumber for the two interaction waves is consistent with the predictions of  (\ref{eq:TT2}).

The global analysis has done a remarkable job of capturing the same key structures observed in experiment. Critically, it has demonstrated that it is indeed possible to describe non-linear KH-shock interactions using a linear model, provided the shock-cell structure is included in the mean flow: the global mode has been shown to correctly capture the non-linear mechanisms underpinning the production of both upstream- and downstream-travelling waves when coherent turbulent structures are convected through a network of shock cells.  The analysis has, furthermore, provided some clarification of the nature of the waves that result from this non-linear interaction. The upstream-travelling wave is the guided jet mode first identified by \cite{TamHu} as a subsonic instability wave. The downstream wave has not previously been discussed in the context of jet screech, but as will be demonstrated it is simply the downstream-travelling form of the same subsonic wave identified by Tam and Hu, which has recently been the subject of extensive investigation by \cite{TowneetalJFM2017} and \cite{schmidt2017wavepackets}. We turn to a local, rather than global, analysis to elucidate the nature of this wave.

\begin{figure}
\centering
\includegraphics[width=1\textwidth]{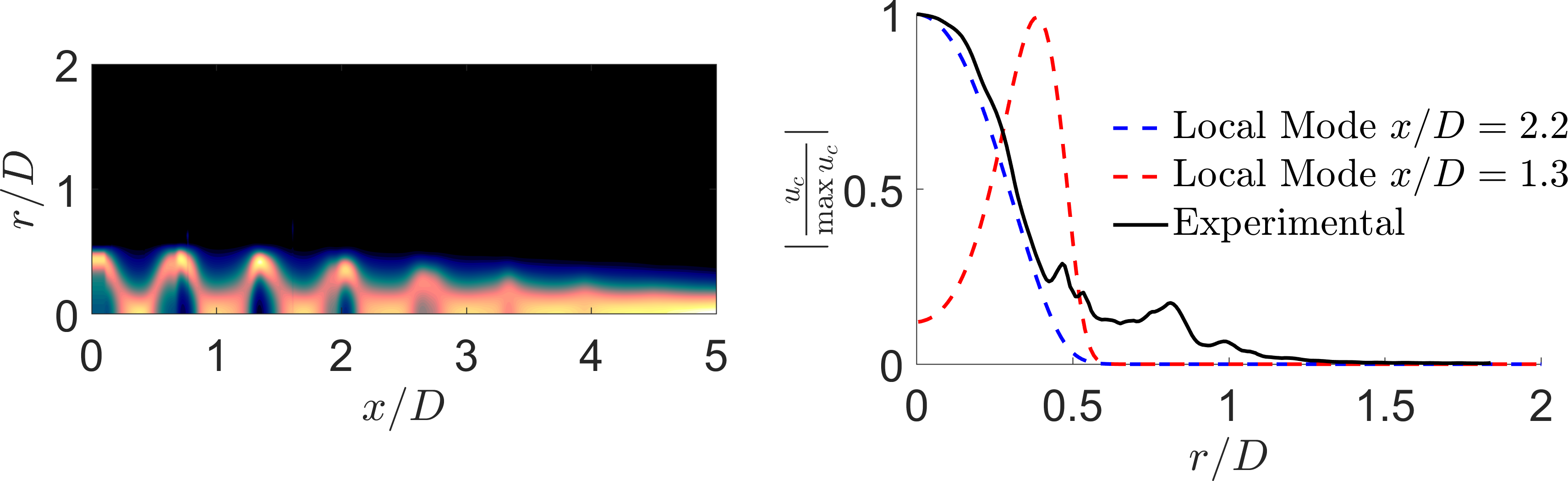}
\caption{Left) Eigenfunctions of the $k^+_{t}$ wave as a function of streamwise position, obtained via local linear stability analysis. Right) Comparison of eigenfunctions at two selected axial locations with profile extracted from figure \ref{fig:NPR210wamp}. Results for $\textit{NPR} = 2.10$ jet.}
\label{fig:blob_r}
\end{figure}

Figure \ref{fig:blob_r} presents radial eigenfunctions of the downstream-propagating subsonic instability wave as a function of axial position; these eigenfunctions have been produced via a local analysis on the experimentally-derived mean flow at each axial location for the screech frequency $St=0.67$. In the region of the flow where the shocks are strongest, both the eigenfunction and the wavenumber (omitted for brevity) of this duct-like wave are significantly modulated by the shocks. Despite this modulation, the wave retains its duct-like character: bounded by the shear layer of the jet. In between the shocks, and further downstream, the wave has the typical structure associated with a $k^+_{t}$ mode of the first radial order, as shown in detail in \cite{TowneetalJFM2017}. Immediately downstream of the shocks however, the wavenumber of the wave significantly increases, and the peak amplitude shifts from the centreline to almost the edge of the shear layer. Considering the overall characteristics of the shock-cell structure, it is possible that high pressure regions oppose the radial support of these waves in the core, such that the peak is moved to regions of moderate pressure. The eigenfunctions at two selected axial locations are compared with the profile extracted from the experimentally derived wavenumber spectrum. At $x/D = 1.3$, the flow is immediately downstream of a shock, and the structure is markedly different to that expected for a $k^+_{t}$ wave. At $x/D = 2.2$, which is close to where the amplitude of this wave peaks according to figure \ref{fig:NPR210wamp}, the experiment and stability analysis match well up to $r/D \approx 0.5$. The wave is not perfectly trapped by the shear layer in the experimental data, and decays more slowly for $r/D > 0.5$. 

With the nature of the three waves now established, we can make an overall statement regarding their behaviours. A non-linear interaction of the KH wavepacket with the shock cells of the jets produces both upstream- and downstream-travelling waves, as predicted by the model of \cite{tam1982shock}; this is the first direct experimental validation of the theory, and likewise the first demonstration that a linear global analysis can capture this non-linear interaction once the shock-cell structure is included in the mean flow. While the model of \cite{tam1982shock} predicted that these waves would exist, it is the mean flow that dictates their nature. In essence, the KH-shock interaction can be thought of as a forcing term, with the response to that forcing dictated by the structure of the mean flow. The mean flow supports upstream- and downstream-travelling waves with distinct radial structures \citep{TamHu}: the $k^-_\textit{th}$ wave has support both inside and outside the jet, while the $k^+_\textit{t}$ wave is (in a vortex-sheet analysis at least) entirely confined to the core of the jet. Thus we now have a full explanation for the results presented here: a ``forcing'' via the mechanism of \cite{tam1982shock}, and a response in the form of waves predicted by \cite{TamHu}. In the following section of the paper, we consider the further interaction between these three waves.

\subsection{Wave superposition in screeching jets}
\begin{figure}
\centering
\includegraphics[width=1\textwidth]{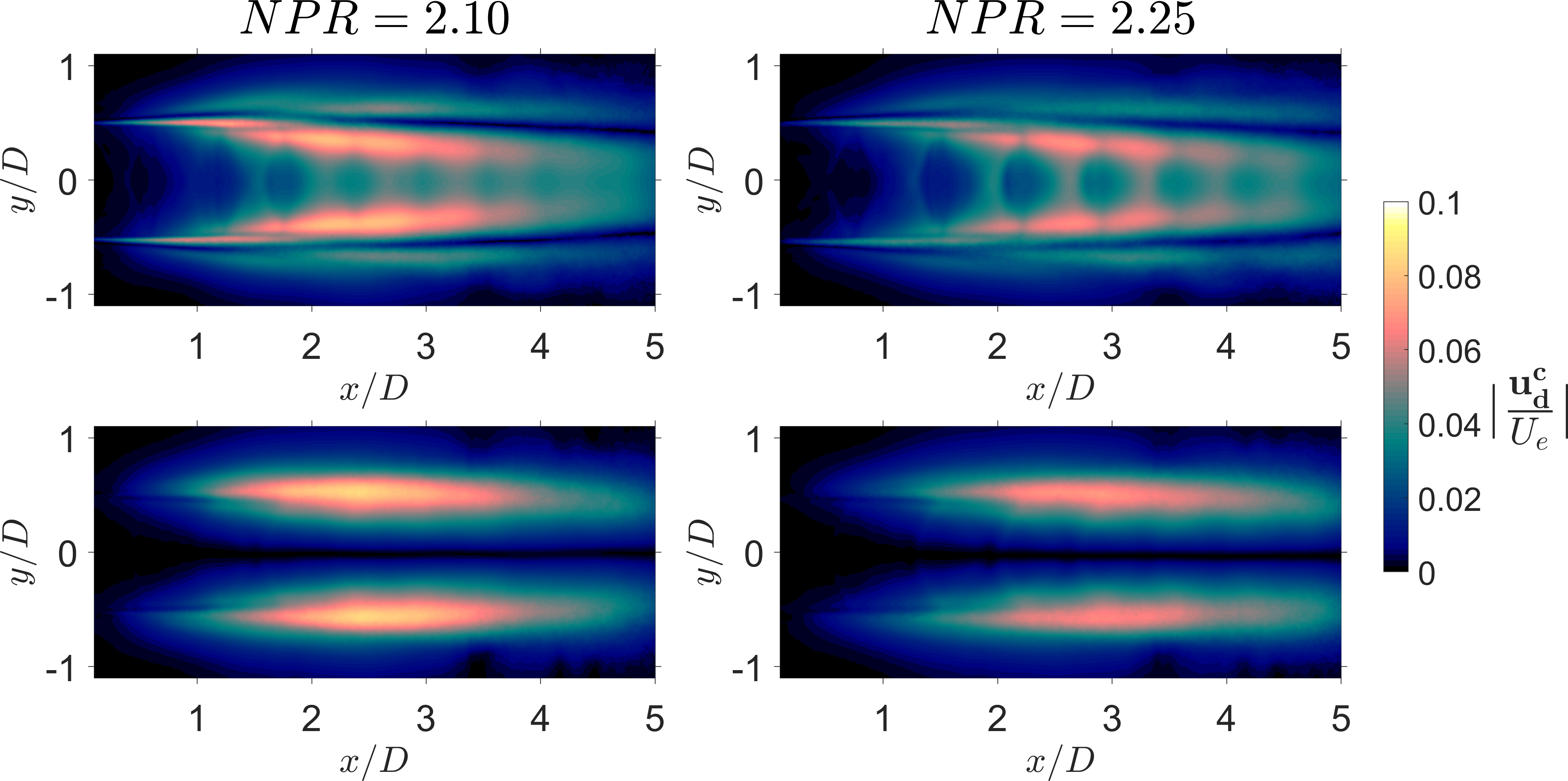}
\caption{Magnitude of coherent fluctuations with positive phase velocity for the two jets where the screech mode is an m = 0 azimuthal mode. Upper) Axial velocity fluctuations. Lower) Transverse velocity fluctuations.}
\label{fig:Down1}
\end{figure}

\begin{figure}
\centering
\includegraphics[width=1\textwidth]{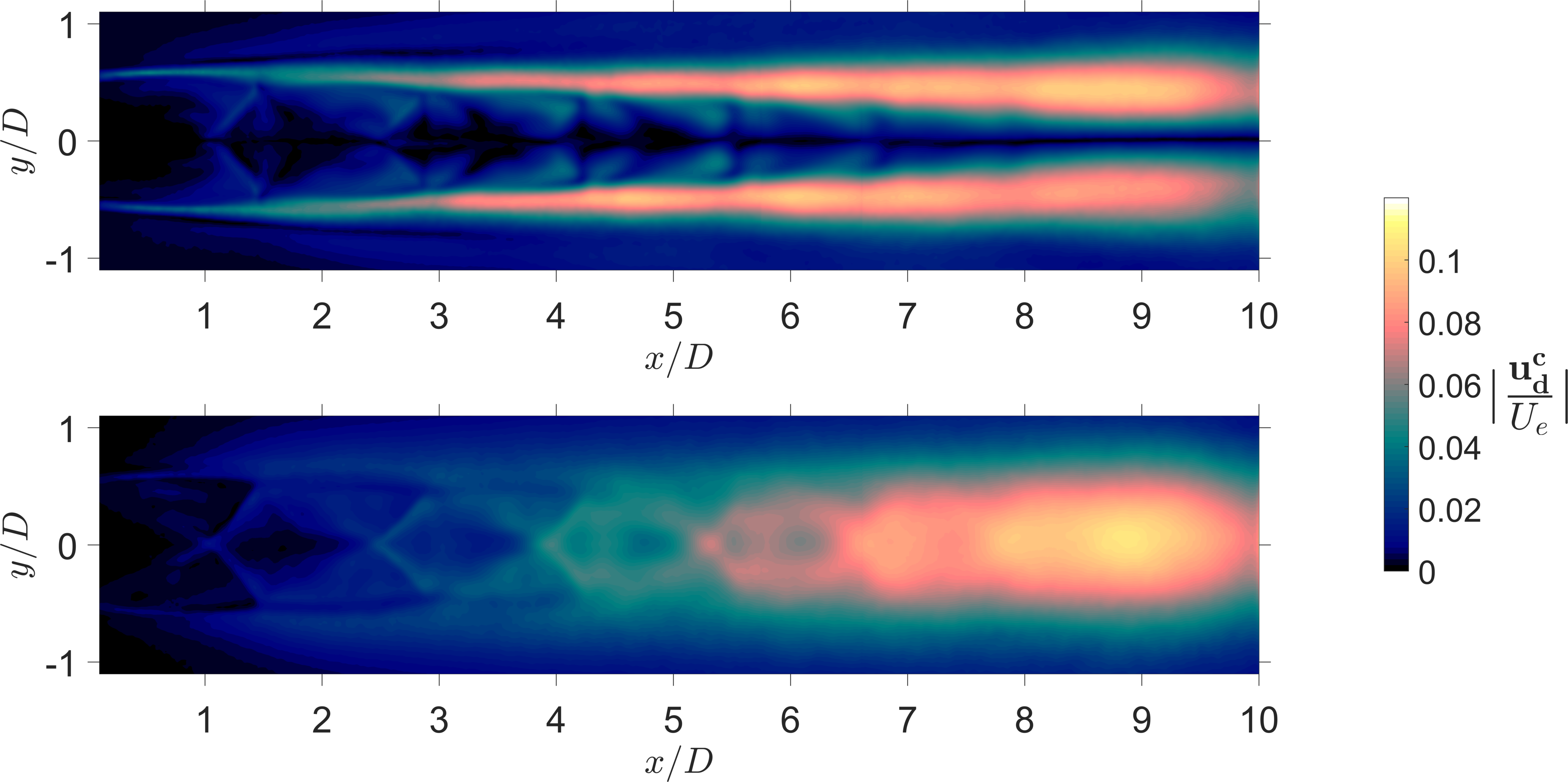}
\caption{Magnitude of coherent fluctuations with positive phase velocity for jet where the screech mode is an m = 1 azimuthal mode. Upper) Axial velocity fluctuations. Lower) Transverse velocity fluctuations.}
\label{fig:Down2}
\end{figure}

As alluded to in the introduction of the paper, and demonstrated in figure \ref{fig:Coh1}, velocity fluctuations in screeching jets are strongly spatially modulated. In this section we consider the source of this modulation. It is already well recognized that the standing wave set up between the downstream-travelling $k^+_\textit{kh}$ waves and the upstream-travelling $k^-_\textit{th}$ waves results in a periodic modulation of the velocity fluctuations. The effect of this modulation can be removed via the application of a high-pass wavenumber filter with a cut-off at $k_x = 0$; all remaining fluctuations are associated with downstream-travelling waves. The results of this filtering are shown in figures \ref{fig:Down1} and \ref{fig:Down2}; all periodic modulation of velocity fluctuations outside of the jet lipline have been removed, but all three jets still exhibit periodic modulation within the jet core. For the $m=0$ screech cases, there is a periodic oscillation of axial velocity fluctuation amplitude in the jet core, but relatively little modulation of the transverse velocity component. For the helical screech mode, there remains significant modulation of velocity in both the axial velocity (primarily on the high-speed side of the shear layer) and the transverse velocity (at the centreline). 

\begin{figure}
\centering
\includegraphics[width=1\textwidth]{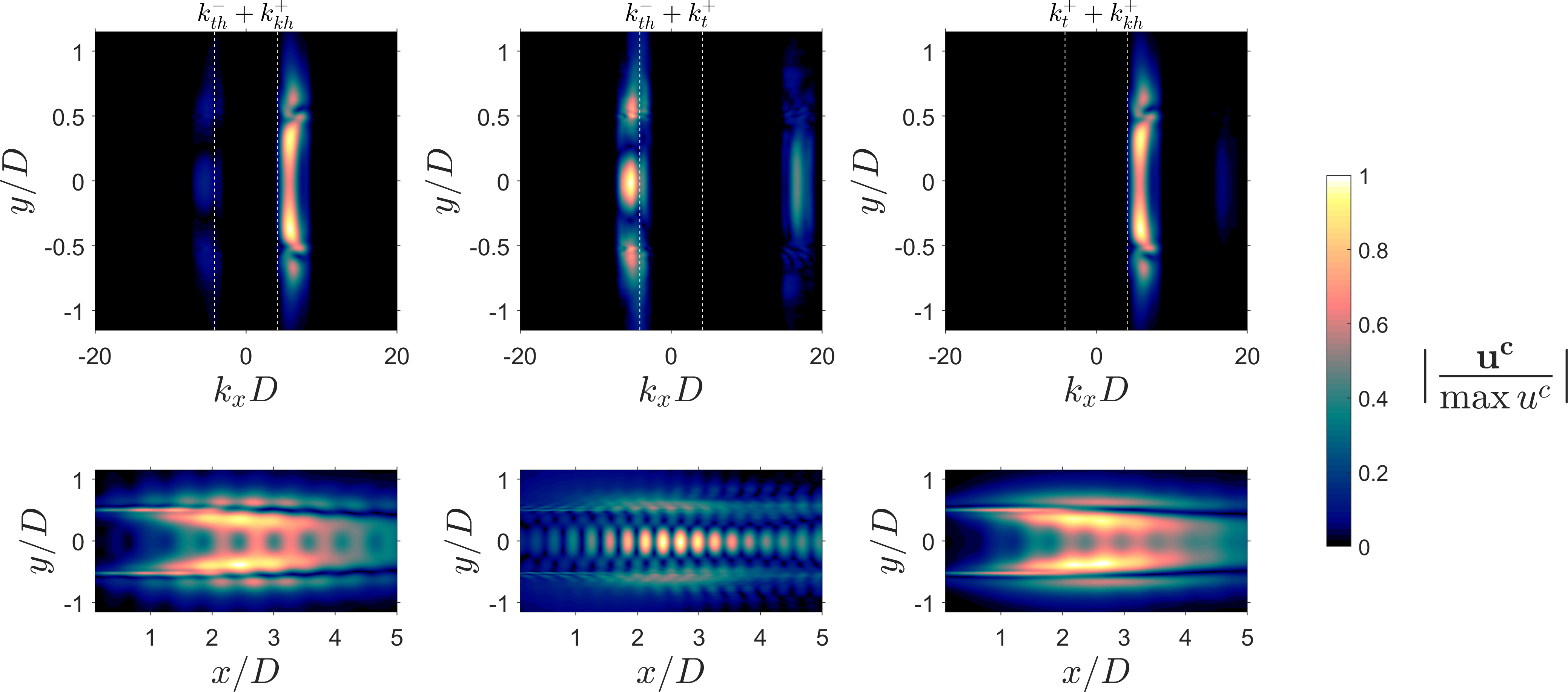}
\caption{Interaction between pairs of waves for  $\textit{NPR} = 2.10$ jet, represented through streamwise velocity fluctuations. Top) Wavenumber spectra. Bottom) Velocity fluctuation amplitude (normalized).}
\label{fig:WP210}
\end{figure}

\begin{figure}
\centering
\includegraphics[width=1\textwidth]{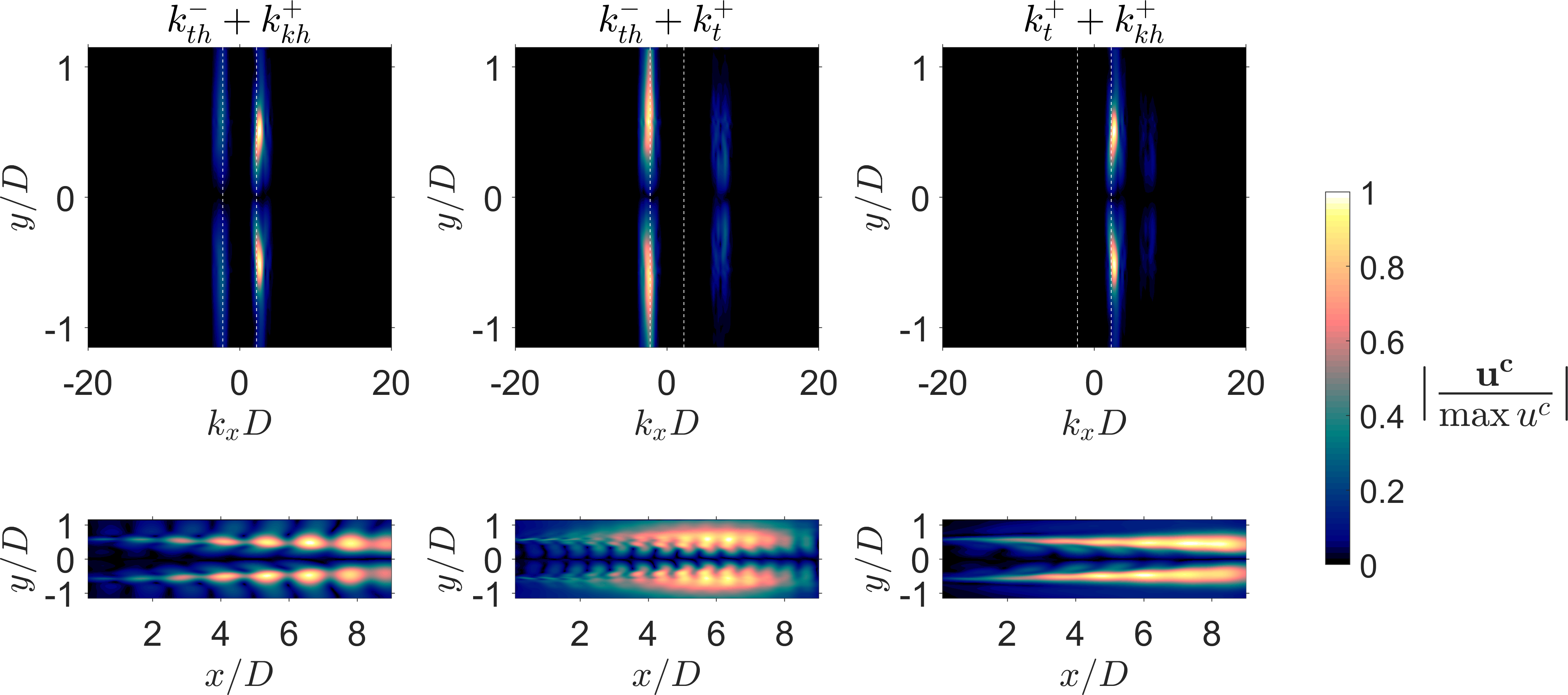}
\caption{Interaction between pairs of waves for $\textit{NPR} = 3.40$ jet, represented through streamwise velocity fluctuations. Top) Wavenumber spectra. Bottom) Velocity fluctuation amplitude (normalized).}
\label{fig:WP340}
\end{figure}

Given the restriction of this modulation to the core of the jet, the obvious candidate is a superposition of the $k^+_{kh}$ wave with the trapped $k^+_{t}$ waves. To examine superposition between all three waves, dual-peak bandpass filters are applied to each of the three possible wave-wave combinations, with both the filtered-wavenumber spectra and resultant amplitude plots presented in figures \ref{fig:WP210} and \ref{fig:WP340}. Again, it must be emphasized that information is lost in this filtering process, and the results should only be considered a qualitative indication. As such, amplitudes have been normalized by the maximum amplitude in each case for clarity, though the fluctuations resulting from the superposition of the $k^-_\textit{th}$ and $k^+_{t}$ waves is an order of magnitude weaker than all others. As expected, the superposition of the $k^-_\textit{th}$ and $k^+_\textit{kh}$ produces the familiar standing-wave pattern in the nearfield of the jet, but is also responsible for significant modulations within the core of the jet. The $k^-_\textit{th}$ and $k^+_{t}$ superposition is highly periodic and confined to the core of the jet, but is very weak and contributes minimally to the overall fluctuations. The $k^+_{t}$ and $k^+_\textit{kh}$ wave superposition, while weaker than that of the $k^-_\textit{th}$ and $k^+_\textit{kh}$, still results in a significant spatial modulation of velocity fluctuation. As the $k^+_{t}$ is confined within the lipline of the jet, so too is the spatial modulation of velocity; outside the jet it is only the superposition of the $k^-_\textit{th}$ and $k^+_\textit{kh}$ waves that produces spatial modulation.
In the flows we consider here, both the $k^-_\textit{th}$ and $k^+_{t}$ waves result from a non-linear interaction between the $k^+_\textit{kh}$ wavepacket and the shocks. The superposition of the $k^+_\textit{kh}$ wave with these two new waves thus produces a spatial modulation whose wavelength closely matches the shock spacing in the flow, but whose nodes and antinodes are not necessarily aligned with the shock locations. 

\begin{figure}
\centering
\includegraphics[width=1\textwidth]{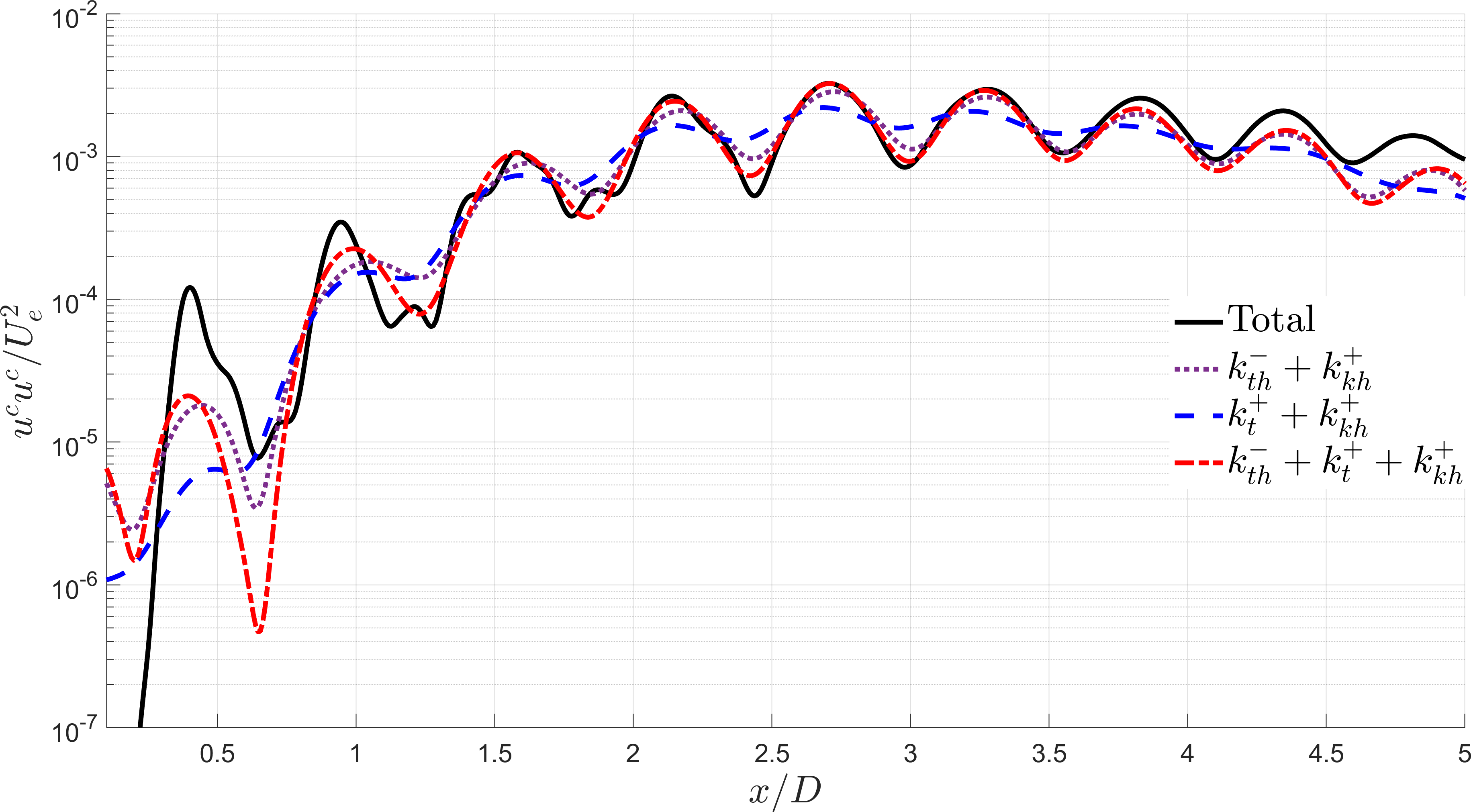}
\caption{Centreline streamwise velocity fluctuation profiles for filtered data subsets of $\textit{NPR} = 2.10$ jet}
\label{fig:WPP210}
\end{figure}

Plots of centreline fluctuation amplitude are presented in figure \ref{fig:WPP210}. Even on the centreline, the majority of the modulation is evidently derived from the superposition of the $k^-_\textit{th}$ and $k^+_\textit{kh}$ waves, with the $k^+_{t}$ and $k^+_\textit{kh}$ providing a small but non-negligible contribution. When the superposition of all three waves is considered, the modulation in the range $2 \leq x/D \leq 4$ is reproduced almost perfectly; outside this range the growth and decay requires the inclusion of lower wavenumbers than those admitted by the filter. Thus the spatial modulation observed in the coherent velocity fluctuations in this jet can be described entirely in terms of the superposition of the three travelling waves identified in this work.

\section{Conclusion}
\cite{tam1982shock} first suggested that the interaction of the KH wavepacket with the shocks in a supersonic jet would produce both waves with both positive and negative phase velocity. Here we have provided the first evidence for these waves in the velocity field of screeching supersonic jets across a range of operating conditions. Further, we have demonstrated that the radial structure of these waves is dictated by the base flow, in accordance with the waves predicted by \cite{TamHu} via a vortex-sheet model. The upstream-travelling wave has support both inside and outside the jet, while the downstream-travelling wave is confined within the core of the jet. This downstream wave is the ``trapped'' wave described in \cite{TowneetalJFM2017}, that treats the jet like a soft-walled duct.

We have also demonstrated that the non-linear wave interaction can be captured by a linear global analysis performed on the experimentally-derived mean flow, thanks to the presence of the shock-cell structure in the mean flow. Despite the many non-linear mechanisms active in a screeching jet, the linear analysis predicts the correct screech tone to within $ \approx 1 \%$, and the same three wave structures observed in experiment are likewise evident in the global mode. This suggests that, aside from the non-linear wave interaction that drives the upstream- and downstream-travelling waves, the remaining frequency-selection mechanisms (propagation characteristics of the $k^+_\textit{kh}$ and $k^-_\textit{th}$ waves, receptivity in the nozzle plane) are linear.

Finally, the superposition of these three wave structures and the spatially-modulated velocity fluctuations that result were considered. We have demonstrated that almost all of the periodic modulation of coherent velocity observed in these screeching jets can be explained by the superposition of these three waves. While two of these waves were produced from an interaction between the KH wavepacket and the shock structures within the jet, we see little evidence that the shocks directly modulate the coherent KH wavepacket. 

\section{Acknowledgements}
Daniel Edgington-Mitchell's contribution to this work was funded by the Australian Research Council under the Discovery Project scheme via DP190102220.

\bibliographystyle{jfm}
\bibliography{JFM_Screech_Wave_structures}

\end{document}